\documentclass[%
 amsmath,amssymb,
 aps,
 pra,
floatfix,longbibliography
]{revtex4-1}
\pdfoutput=1
\usepackage{graphicx}
\usepackage{dcolumn}
\usepackage{bm}
\usepackage{subfigure}
\usepackage{overpic}
\usepackage[export]{adjustbox}

\usepackage{color}

\AtBeginDocument{%
 \newwrite\bibnotes
 \def\bibnotesext{Notes.bib}
 \immediate\openout\bibnotes=\jobname\bibnotesext
 \immediate\write\bibnotes{@CONTROL{REVTEX41Control}}
 \immediate\write\bibnotes{@CONTROL{%
   apsrev41Control,author="08",editor="1",pages="1",title="0",year=?1?}}
 \if@filesw
 \immediate\write\@auxout{\string\citation{apsrev41Control}}%
 \fi
}%

\begin{document}
\newcommand{\reva}[1]{{\color{black}{#1}}}
\newcommand{\revb}[1]{{\color{black}{#1}}}
\newcommand{\revc}[1]{{\color{black}{#1}}}


\title{Secondary atomization of liquid columns in compressible crossflows}

\author{Daniel P. Garrick}
\altaffiliation[Now at ]{Theta Solutions LLC, Atascadero, CA 93422, USA and A.L. Rae Centre for Genetics \& Breeding, School of Agriculture and Environment, Massey University, New Zealand.}
\author{Wyatt A. Hagen}
\altaffiliation[Now at ]{University of Illinois, Urbana, IL 61801, USA}
\author{Jonathan D. Regele} \email{jregele@lanl.gov}
\altaffiliation[Now at ]{Los Alamos National Laboratory, Los Alamos, NM 87545, USA.}
\affiliation{Department of Aerospace Engineering, Iowa State University, Ames, IA 50011, USA}%





\begin{abstract}
The secondary atomization of liquid droplets is a common physical phenomenon in many industrial and engineering applications. Atomization in high speed compressible flows is less well understood than its more frequently studied low Mach number counterpart. The key to understanding the mechanisms of secondary atomization is examination of the breakup characteristics and droplet trajectories across a range of physical conditions. In this study, a planar shock wave impacting a cylindrical water column ($\rho_l=1000 \rm~kg/m^3$) is simulated for a range of Weber numbers ranging three orders of magnitude ($\sim 10^0-10^3$). Four different incident shock speeds are simulated ($M_s=1.47, 2, 2.5, 3$) which induce subsonic, transonic, and supersonic crossflow across the column. The flowfield is solved using a compressible multicomponent Navier-Stokes solver with capillary forces. Fluid immiscibility is maintained with an interface sharpening scheme. Overall, a diverse range of complex interface dynamics are captured across the range of physical conditions studied. Additionally, while the unsteady drag coefficient of the liquid column shows a dependence on the Weber number using the undeformed diameter, calculations using the deformed diameter significantly reduce the dependence, particularly for the supersonic cases, with implications for subgrid droplet modeling in atomization simulations. A preliminary under-resolved three-dimensional simulation of droplet breakup shows reasonable agreement with experimental data, indicating the potential of the numerical approach for future investigations.


\end{abstract}

\maketitle


\section{Introduction}
Liquid atomization is an important physical process in a wide variety of applications ranging from manufacturing (including 3D printing) to drug delivery and fuel sprays.
The process of liquid breakup has a strong dependence on the Weber number which relates the inertial force to the surface tension.
As a large quantity of atomization applications occur in low Mach number flow regimes, significant numerical modeling effort has focused on incompressible schemes~\cite{Gorokhovski2008}. \reva{State of the art secondary atomisation modeling in the compressible flow regime has largely focused on the early stages of the breakup process and/or higher Weber numbers where the effects of surface tension are assumed to be negligible and are not considered \cite{Meng2018,Liu2018,Xiang2017}.} 
Meanwhile, technical challenges involving supersonic combustion ramjets (scramjets) has identified a need for greater understanding of the penetration, mixing, and atomization of liquid jets injected into high-speed compressible crossflows~\cite{Lee2015}.

Liquid jet atomization consists of primary and secondary breakup.
The former consists of the bulk liquid transforming into smaller jets, sheets, and droplets.
Secondary breakup consists of liquid droplets or ligaments undergoing further deformation and breakup and has generally been classified into vibrational, bag, multi-mode (or bag-and-stamen), sheet-thinning, and catastrophic regimes according to the Weber number~\cite{Guildenbecher2009,Pilch1987,Hsiang1992,Faeth1995}. \reva{However, Theofanous et al.~\cite{Theofanous2004} examined droplet breakup in highly rarefied supersonic flow conditions and instead proposed classification of the breakup into two primary criticalities, Rayleigh-Taylor piercing (RTP) and shear-induced entrainment (SIE). The defining feature of RTP is the penetration of the droplet by the gas while SIE is demarcated by a breakup process involving a peeling of the outer surface of the droplet~\cite{Theofanous2012}.  As noted by Guildenbecher et al.~\cite{Guildenbecher2009}, this departure from the traditional breakup morphology suggests more investigation of the topic is needed. Moreover, several researchers have pointed out a dependence of the breakup behavior on the density ratio~\cite{Jalaal2014,Han2001} which is important in the context of high speed flows with varying post-shock gas densities and significant compressibility effects.}
Simulating the entire atomization process requires extremely high resolution due to the multiscale nature of the features involved.
This is especially problematic at high Reynolds and Weber numbers where resolving the boundary layer on the droplet surface and becomes difficult and large numbers of small droplets can be generated.
 Subgrid droplet models can relax the computational complexity and have been used to simulate liquid jet injection in supersonic crossflows~\cite{Kim2012,Liu2016}.
However they generally utilize steady-state empirical relations for the drag coefficient of solid spherical particles as a function of the particle Reynolds number to calculate drop trajectories~\cite{crowe2011multiphase}.

To better understand the behavior of deforming droplets in crossflows and the secondary atomization process in general, various experimental and numerical studies have been performed and were recently reviewed by Guildenbecher et al~\cite{Guildenbecher2009}.
With respect to the drag coefficient, Kim et al.~\cite{Kim1998} found that the effects of the initial relative velocity and large relative acceleration or deceleration are significant when predicting rectilinear motion of spherical particles in crossflows.
Experiments by Temkin and Mehta~\cite{temkin1982} showed that the unsteady drag is always larger in decelerating or smaller in accelerating flows than the steady state value.
Wadhwa et al.~\cite{Wadhwa2007} coupled a compressible gas phase solver with an incompressible liquid phase solver and found for axisymmetric conditions the droplet Weber number affects the drag coefficient of a drop traveling at high speeds and placed in quiescent air.
 Finally, the unsteady nature of the flow as well as the scales (both temporal and spatial) involved in droplet breakup means experimentally measuring the local drop and ambient flow fields during secondary atomization is incredibly challenging~\cite{Guildenbecher2009}.
Therefore, numerical simulations are a valuable tool for providing important physical insight in such conditions.
While some experimental~\cite{Theofanous2004} and numerical~\cite{Chang2013} investigations exist on the interface dynamics and breakup behavior of liquid droplets at a handful of supersonic flow conditions and Weber numbers, the secondary atomization process across a diverse range of physical conditions has not yet been investigated thoroughly.

Experimental investigation of liquid columns (as opposed to spherical droplets) allows for easier visualization of the wave structures~\cite{Igra2001,Sembian2016}, although difficulties remain in visualizing the later stages of the breakup process.
The deformation behavior of the two-dimensional liquid columns have also been found to follow similar trends as that of three-dimensional spherical droplets~\cite{Igra2002,Igra2010}.
Numerous researchers have simulated the two-dimensional shock-column interaction, commonly as a test case for compressible multicomponent flow solvers~\cite{Igra2010,Meng2014,Shukla2010,Shukla2014,Terashima2009,Terashima2010,Chen2008,Nonomura2014}.
 Notable examples include the work of Terashima and Tryggvason~\cite{Terashima2009} who simulated the entire evolution of the column breakup, while Meng and Colonius~\cite{Meng2014} and Chen~\cite{Chen2008} examined the sheet-thinning process and evaluated column trajectories and drag coefficients.
However, such studies focused on the early stages of breakup and neglected the effects of both surface tension and molecular viscosity.
As a result, questions remain as to the breakup process of a liquid column when accounting for molecular viscosity and surface tension effects and especially in the context of supersonic flows.
Fortunately, the cylindrical geometry of the water column can be efficiently modeled using a two-dimensional domain providing faster turnaround times compared to full three-dimensional simulations.
This allows a wider range of physical conditions to be efficiently examined where for similar reasons axisymmetric domains and/or lower gas-liquid density ratios have been employed in incompressible studies~\cite{Strotos2016,Han1999,Han2001}.

Garrick et al.~\cite{Garrick2016} performed a preliminary study of secondary atomization without molecular viscosity effects and while using a non-conservative interface sharpening scheme.
Several simulations of water column-shock interactions were performed including an $M_s=1.47$ shock with comparisons to experiment and an $M_s=3$ shock with and without surface tension.
These simulations considered the early stages of breakup and successfully highlighted the effects of surface tension on the dynamics of the gas-liquid interface.
The dependence of the breakup behavior on the Weber number for $\mathrm{We}=5-100$ was also examined with an array of $M_s=1.39$ ($M=0.5$ crossflow) shock-column simulations.
The liquid-gas density ratio was set to $\rho_l/\rho_g=10$ to reduce computational effort.
Garrick et al.~\cite{Garrick2016a} extended the numerical method to account for molecular viscosity and non-uniform grids and replaced the non-conservative interface sharpening scheme with a conservative reconstruction based interface sharpening scheme.
That approach was then applied to simulate primary and secondary atomization in high speed crossflow.
The present work applies the same approach to a wider range of secondary atomization conditions for a two-dimensional liquid column with a high density ($\rho_l=1000  \mbox{ kg}/\mbox{m}^3$).
This should provide a first order estimate of the three-dimensional behavior but with the benefit of a significantly reduced computational cost. 

%
%

\reva{To gain a better understanding of the secondary atomization process in high speed flows, the present work simulates shock-column interactions at various Weber and incident shock Mach numbers to examine the combined effects of surface tension and compressibility on the breakup process across a broad range of physical conditions.
This involves detailed two-dimensional simulations of column breakup in high speed compressible flows while accounting for capillary and viscous forces and utilizing an interface sharpening scheme to maintain the fluid immiscibility condition and prevent unphysical numerical smearing of the interface.
Particular focus is made on the breakup process and drag coefficient of the droplets over time.}
The two-dimensional nature of the study is motivated by the focus on a broad range of physical conditions which would be otherwise cost prohibitive to simulate in three dimensions.
This follows prior studies which utilized two-dimensional or axisymmetric domains (see~\cite{Meng2014,Han1999,Han2001,Chen2008,Chen2008a,Igra2001a}) and is also motivated by experimental observations of qualitatively similar breakup characteristics for two-dimensional liquid columns and three-dimensional spherical droplets~\cite{Igra2001,Igra2002}.

The paper is organized as follows.
Section~\ref{sec:modeling3} describes the mathematical model and non-dimensionalization.
Section~\ref{sec:num3} describes the numerical approach while the problem statement is reviewed in Section~\ref{sec:problem3}.
\reva{Section~\ref{sec:results3} presents a two-dimensional investigation of the breakup process and drag coefficient of a liquid column across a range of Weber and incident shock Mach numbers.}
\revc{This is followed with a three-dimensional droplet breakup simulation in Section~\ref{sec:3dsim} and conclusions in Section~\ref{sec:conclusions3}.}
\section{Mathematical model}\label{sec:modeling3}
The present work utilizes the approach of Garrick et al.~\cite{Garrick2016,Garrick2016a} for solving the flowfield.
A non-dimensional form of the quasi-conservative five equation model of Allaire~\cite{Allaire2002} is employed with capillary and molecular viscosity terms.
As such, the compressible multicomponent Navier-Stokes equations govern the flowfield~\cite{Perigaud2005}:
\begin{subequations}
	\begin{align}
	&\frac{\partial \rho_1 \phi_1}{\partial t} + \nabla \cdot (\rho_1 \phi_1 \mathbf{u}) = 0, \label{eqn:finala} \\
	&\frac{\partial \rho_2 \phi_2}{\partial t} + \nabla \cdot (\rho_2 \phi_2 \mathbf{u}) = 0, \\
	&\frac{\partial \rho \mathbf{u}}{\partial t} + \nabla \cdot (\rho \mathbf{u} \mathbf{u} + p \tilde{I}) =  \frac{1}{\mathrm{Re_a}} \nabla \cdot \boldsymbol{\tau} + \frac{1}{\mathrm{We_a}} \kappa \nabla \phi_1, \\
	&\frac{\partial E}{\partial t} + \nabla \cdot \left( ( E + p) \mathbf{u} \right) = \frac{1}{\mathrm{Re_a}} \nabla \cdot (\boldsymbol{\tau} \cdot \mathbf{u}) + \frac{1}{\mathrm{We_a}} \kappa \nabla \phi_1 \cdot \mathbf{u}, \\
	&\frac{\partial \phi_1}{\partial t} + \mathbf{u} \cdot \nabla \phi_1 = 0, \label{eqn:finale} 
	\end{align}
\end{subequations}
where $\rho_1 \phi_1$, $\rho_2 \phi_2$, and $\rho$ are the liquid, gas, and total densities, $\mathbf{u}=(u,v)^T$ is the velocity, $\phi_1$ is the liquid volume fraction, $p$ is the pressure, $\mathrm{We_a}$ and $\mathrm{Re_a}$ are the acoustic Weber and Reynolds numbers, respectively, $\kappa$ is the interface curvature, and $E$ is the total energy
\begin{equation}
E = \rho e + \frac{1}{2}\rho \mathbf{u} \cdot \mathbf{u}
\end{equation}
where $e$ is the specific internal energy.

\revb{The model is non-dimensionalized using the rules in Table~\ref{tab:nondim2} where primes indicate dimensional quantities and the subscript `$0$' refers to a chosen reference state. The dimensional distance $l^\prime_0$ is chosen as the droplet diameter.
This results in the viscous and capillary forces being scaled by acoustic Reynolds and Weber numbers:
\begin{align}
\mathrm{Re_a} &=\frac{\rho^\prime_0 a^\prime_0 l^\prime_0}{\mu^\prime_0} \\
\mathrm{We_a} &=\frac{\rho^\prime_0 a^{\prime 2}_0 l^\prime_0}{\sigma^\prime_0} 
\end{align}
where $\mu^\prime_0$ and $\sigma^\prime_0$ are the reference dimensional viscosity and surface tension coefficients, respectively.}

The viscous stress tensor $\boldsymbol{\tau}$ is given with the non-dimensional mixture viscosity $\mu$:
\begin{equation}
\boldsymbol{\tau} = 2 \mu \left( \mathbf{D} - \frac{1}{3} (\nabla \cdot \mathbf{u}) \mathbf{I} \right)\label{eqn:tau}
\end{equation} 
where $\mathbf{D}$ is the deformation rate tensor
\begin{equation}
\mathbf{D} = \frac{1}{2} \left( \nabla \mathbf{u} + ( \nabla \mathbf{u} )^T \right).
\end{equation}
The fluid components are considered immiscible and the liquid and gas volume fraction functions ($\phi_1$ and $\phi_2$ respectively) are used to capture the fluid interface.
Mass is discretely conserved for each phase via individual mass conservation equations.
Surface tension is implemented as a volume force as in the CSF model~\cite{Alamos1992} with terms in both the momentum and energy equations~\cite{Perigaud2005}.
While a conservative form of the \reva{surface tension term} exists~\cite{Gueyffier1999}, the present model utilizes the non-conservative form which enables flexible treatment of the curvature term $\kappa$ and its accuracy.

\begin{table}\centering
	\caption{Non-dimensional rules used in the model.\label{tab:nondim2}}
	\begin{tabular}{cc}
		Parameter & Rule \\
		\hline 
		Position & $x=x^\prime/l^\prime_0$ \\ 
		Time & $t=t^\prime a^\prime_0/l^\prime_0$ \\ 
		Velocity & $u=u^\prime/a^\prime_0$ \\ 
		Density & $\rho=\rho^\prime/\rho'_0$ \\ 
		Pressure & $p=p^\prime/\rho^\prime_0 a^{\prime 2}_0$ \\ 
		Total Energy & $E=E^\prime/\rho^\prime_0 a^{\prime 2}_0$ \\ 
		Curvature & $\kappa=\kappa^\prime l^\prime_0$ \\ 
		Surface tension coefficient & $\sigma = \frac{1}{\mathrm{We_a}} = \frac{\sigma^\prime_0}{\rho^\prime_0 a_0^{\prime 2} l^\prime_0}$ \\
		Viscosity & $\mu = \frac{1}{\mathrm{Re_a}} = \frac{\mu^\prime_0}{\rho^\prime_0 a_0^\prime  l^\prime_0}$ \\
		\hline
	\end{tabular} 
\end{table}

\subsection{Equation of state and mixture rules} 
To close the model, the stiffened gas equation of state (EOS)~\cite{stiffenedgas} is employed to model both the gas and liquid phases. \revb{The stiffened gas equation of state utilises fitting parameters $\gamma$ and $\pi_\infty$ to recreate the sonic speed in various materials based on experimental measurements. In the case of air, $\gamma=1.4$ becomes the specific heat ratio with $\pi_\infty=0$ and the stiffened gas equation of state simplifies to the ideal gas law. For a given simulation containing a liquid ($1$) and gas ($2$), the stiffened gas equation of state fitting parameters are computed at every point within the domain as a function of the volume fraction:
\begin{equation}
\Gamma = \frac{1}{\gamma-1} = \frac{\phi_2}{\gamma_2 - 1} + \frac{\phi_1}{\gamma_1-1}\label{eqn:gam3}
\end{equation}
and
\begin{equation}
\Pi = \frac{\gamma \pi_\infty}{\gamma-1} = \frac{\phi_2 \gamma_2 \pi_{\infty,2}}{\gamma_2 - 1} + \frac{\phi_1 \gamma_1 \pi_{\infty,1}}{\gamma_1-1}.\label{eqn:pi3}
\end{equation}
where $\gamma_1$, $\gamma_2$, $\pi_{\infty,1}$,  and $\pi_{\infty,2}$ are the specific stiffened gas EOS fitting parameters for the liquid ($1$) and gas ($2$). Using the mixture quantities $\Gamma$ and $\Pi$ the total energy becomes
\begin{equation}
E = \Gamma p + \Pi + \frac{1}{2}\rho \mathbf{u} \cdot \mathbf{u}.
\end{equation}

The speed of sound is given by
\begin{equation}
c = \sqrt{ \frac{ \gamma (p + \pi_\infty) }{\rho} }
\end{equation}
where the stiffened gas EOS fitting parameters $\gamma$ and $\pi_\infty$ are computed using the mixture quantities in Eqs.~\ref{eqn:gam3} and~\ref{eqn:pi3}.}
Similar to Coralic and Colonius~\cite{Coralic2014}, the mixture viscosity is determined following Perigaud and Saurel~\cite{Perigaud2005} but written in non-dimensional form for use in Eq.~\ref{eqn:tau}:
\begin{align}
\mu &= \frac{\mu^\prime_1}{ \mu^\prime_0}\phi_1 + \frac{\mu^\prime_2}{ \mu^\prime_0} \phi_2 \notag\\
&= N\phi_1 + \phi_2
\end{align}
where the liquid ($1$) and gas ($2$) viscosities are assumed to remain constant with the gas viscosity used as the reference state $\mu^\prime_0$.
As a result, $\mu^\prime_2/ \mu^\prime_0=1$ and $N=\mu^\prime_1/ \mu^\prime_0$ becomes the liquid to gas viscosity ratio.

\section{Numerical method}\label{sec:num3}

 
The model (Eqs.~\ref{eqn:finala}-\ref{eqn:finale}) is discretized using a finite volume method on a non-uniform two-dimensional Cartesian grid.
The convective fluxes are upwinded using the Harten-Lax-van Leer-Contact (HLLC) approximate Riemann solver originally developed by Toro et al.~\cite{Toro1994,Toro2009} with modifications for surface tension by Garrick et al.~\cite{Garrick2016}.
Following the approach of Johnsen and Colonius~\cite{Johnsen2006}, oscillation free advection of material interfaces is ensured with adaptations to the HLLC for a quasi-conservative form of the volume fraction transport equation.
Viscous terms are implemented following Coralic and Colonius~\cite{Coralic2014}.
Spatial reconstruction to cell faces is performed on the primitive variables using the second order MUSCL scheme with the minmod limiter.
The fluid immiscibility condition is maintained using the $\rho$-THINC interface sharpening procedure~\cite{Garrick2016a} for reconstructing the phasic densities and volume fraction within the interface.
The conserved variables are integrated in time using an explicit third order TVD Runge-Kutta scheme~\cite{Gottlieb1996}.
Interface curvature is calculated via the interface normals ($\kappa=-\nabla \cdot \mathbf{n}$) which are determined using the smoothed interface function of Shukla et al.~\cite{Shukla2010} and second order central differences.
A full description of the numerical method employed and the results of standard validation cases can be found in the work of Garrick et al.~\cite{Garrick2016,Garrick2016a}.

\subsection{Water column attached domain}
Additional computational efficiency is gained by translating the static domain with the $x$ component of the column center of mass.
This requires appropriate modifications to the fluxes via a simplified arbitrary Lagrangian Eulerian (ALE) formulation~\cite{Luo2004}.
The liquid center of mass (and thus the moving grid) velocity $u_c$ is determined via~\cite{Meng2014}:
\begin{equation}
u_c = \frac{\int \rho_1\phi_1 u dV}{\int \rho_1 \phi_1 dV}.\label{eqn:uc}
\end{equation}
The individual control volumes remain static, however, the overall computational domain translates downstream such that the liquid center of mass remains approximately centered throughout the simulation.

\subsection{Drag coefficient}\label{sec:drag}
In the present study the drag coefficient of the liquid is computed following the approach of Meng and Colonius~\cite{Meng2014}:
\begin{equation}
C_d = \frac{m a_c}{\frac{1}{2} \rho_g ( u_g - u_c)^2 d_0} \label{eqn:cd}
\end{equation}
where $d_0$ is the undeformed diameter of the column, $\rho_g$ and $u_g$ are the initial post-shock gas conditions and $u_c$ is the center of mass velocity given by equation~\ref{eqn:uc}.
The acceleration is then computed using finite differences in time~\cite{Meng2014}:
\begin{equation}
a_c = \frac{d}{dt} \frac{\int \rho_1\phi_1 u dV}{\int \rho_1 \phi_1 dV}.\label{eqn:ac}
\end{equation}

\section{Problem statement}\label{sec:problem3}

\begin{figure}\centering
\includegraphics[width=0.5\textwidth]{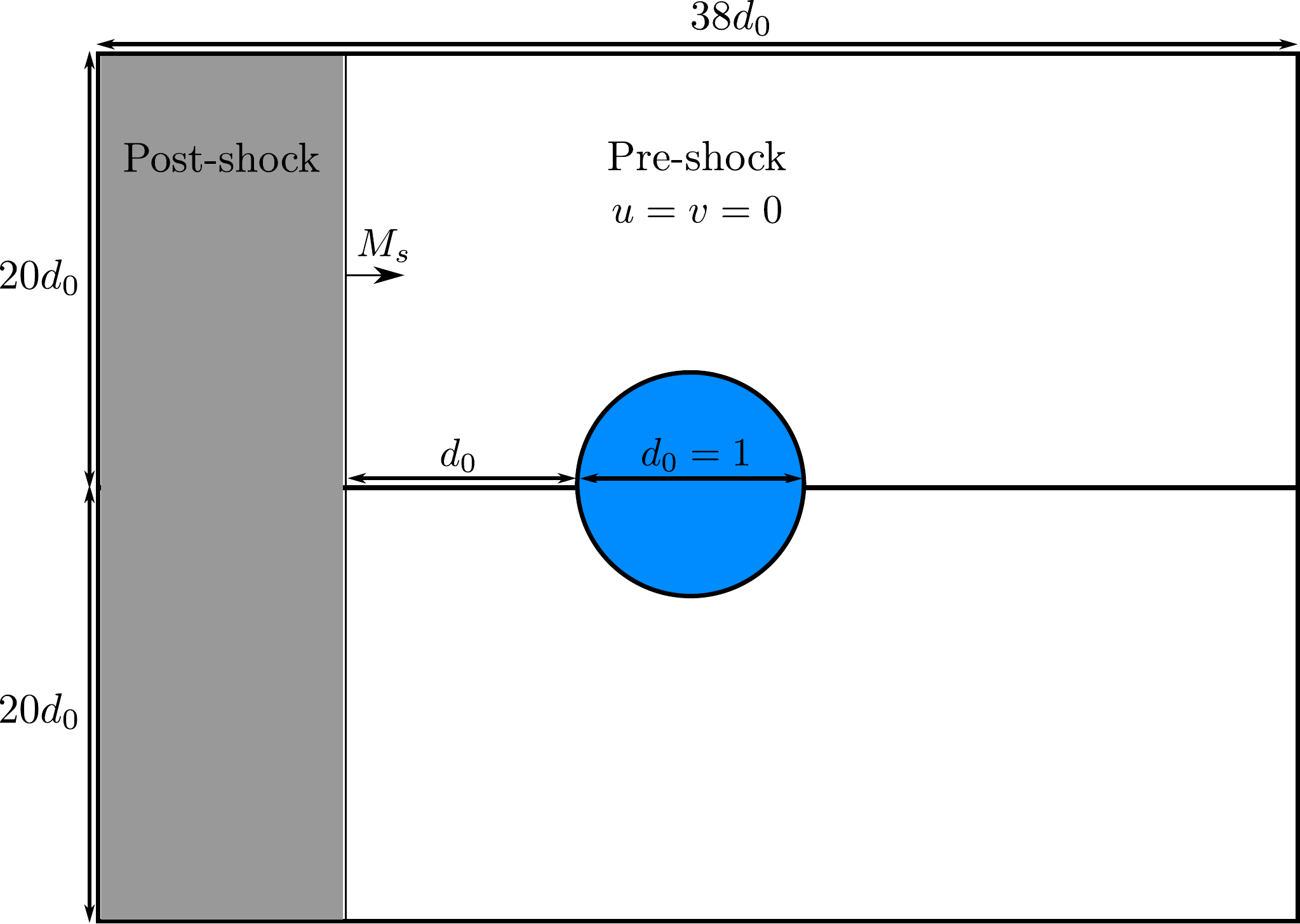} 
\caption{\label{fig:domain}Initial layout of the two-dimensional computational domain.
The liquid column has a unity non-dimensional diameter and is centered at the origin.}
\end{figure}

Standard benchmark cases to verify and validate the shock and interface capturing scheme and the implementation of surface tension were performed by Garrick et al.~\cite{Garrick2016,Garrick2016a}.
For the present simulations, the initial conditions are depicted in Figure~\ref{fig:domain} and correspond to a liquid column ($\rho_l=1000  \mbox{ kg}/\mbox{m}^3$) in air ($\rho_g=1.2 \mbox{ kg}/\mbox{m}^3$) at ambient pressure ($p=101325\mbox{ Pa}$).
The column has unity non-dimensional diameter and is centered at the origin.
Dirichlet and extrapolation conditions are enforced on the upstream and remaining boundaries respectively.
The domain consists of a block of uniform cells in the vicinity of the column corresponding to a resolution of 120 points across the initial column diameter.
Grid stretching to the boundary results in an overall domain of $1579\times1589$ cells.

Simulations are performed for incident shock Mach numbers of $M_s=1.47$, $M_s=2$, $M_s=2.5$, and $M_s=3$. \revb{The incident shock wave is traveling at a speed defined by the incident shock Mach number toward the liquid column which is stationary in ambient air conditions. The Mach number of the induced crossflow for each simulation is determined by first employing the normal shock relations to compute the Mach number and local speed of sound in the gas behind the incident shock. The crossflow Mach number is the ratio of the post-shock (crossflow) gas velocity in the shock moving reference frame to the post-shock speed of sound.}
Passage of these incident shocks over the liquid column induces a crossflow with corresponding Mach numbers of $M=0.58$, $M=0.96$, $M=1.2$, and $M=1.36$, respectively, \revb{which range from subsonic to supersonic speeds. These initial conditions are analagous to experimental shock tube setups whereby pressurized gas is released from a driver section into a driven section such that a shock wave develops and travels down the tube to produce a uniform step change in velocity over droplets inserted into the driven section~\cite{Guildenbecher2009}.}

The \reva{surface tension term} in the momentum and energy conservation equations is scaled by the acoustic Weber number.
To examine the breakup behavior for a range of physical conditions, simulations with $\mathrm{We_a}=1, 5, 10, 20, 50, 100, \mbox{ and } 1000$ were performed for each incident shock speed.
In addition, the breakup behaviors for $\mathrm{We_a}=0.05 \mbox{ and } 0.2$ are considered for the $M_s=3$ incident shock speed.
The acoustic Reynolds number was held constant with a value of $\mathrm{Re_a}=1000$ and a liquid to gas viscosity ratio of $N=\mu_l/\mu_g=45$.
 In the dimensional sense and for a given surface tension coefficient, each acoustic Weber number represents a different column diameter.
Of particular interest is the difference in breakup behavior in subsonic versus supersonic crossflow across the range of Weber numbers.

To quantify the strength of the \reva{surface tension} for each simulation, several Weber numbers are described.
These are the acoustic, crossflow, and effective Weber numbers.
The acoustic Weber number is given in terms of the reference quantities used to non-dimensionalize the system:
\begin{equation}
\mathrm{We_a} =\frac{\rho^\prime_0 a^{\prime 2}_0 d_0^\prime}{\sigma^\prime_0}.
\end{equation}
Meanwhile the crossflow Weber number $\mathrm{We_c}$ is computed using the post-shock crossflow conditions:
\begin{equation}
\mathrm{We_c} =\mathrm{We_a} \rho u^2 \label{eqn:wep}
\end{equation}
where $u$ is the non-dimensional streamwise flow speed and $\rho$ is the non-dimensional density behind the incident shockwave.
The crossflow Reynolds number is similarly estimated by scaling the acoustic Reynolds numbers by the initial post-shock conditions to give $\mathrm{Re}_{1.47}=1430$, $\mathrm{Re}_{2}=4000$, $\mathrm{Re}_{2.5}=7000$, and $\mathrm{Re}_{3}=10290$ for the $M_s=1.47, 2, 2.5$ and $M_s=3$ cases respectively.

Based on the crossflow Reynolds and Weber numbers, these simulations correspond to Ohnesorge numbers ranging from 0.001 to 0.045.
Finally, all simulation times are scaled into their respective non-dimensional characteristic times given by~\cite{Nicholls1969}:
\begin{equation}
t^*= \frac{t u}{D \sqrt{\epsilon}}
\end{equation}
where $u$ is the crossflow velocity and $\epsilon$ is the liquid to gas density ratio using the post-shock conditions. \revb{ The presence of the density ratio in this equation indicates some dependence of the breakup behavior on the local density ratio which varies for each incident shock Mach number as the post-shock gas density varies depending on the strength of the incident shock. In addition, for the simulations with supersonic crossflow a bow shock is generated in front of the liquid column, further compressing the gas. As a result the local gas-liquid density ratio varies considerably for each incident shock Mach number. 
	
	One approach to quantify the compressibility effects is the computation of an effective Weber number which considers the local flow conditions that occur behind the bow shock for the simulations with a supersonic crossflow. This effective Weber number can be computed using the crossflow Mach and Weber numbers and the velocity and density normal shock relations~\cite{Xiao2016}: 
	\begin{equation}
	\mathrm{We_{eff}} = \frac{2+(\gamma-1)M^2}{(\gamma+1)M^2}\mathrm{We_c}.\label{eqn:weeff}
	\end{equation} 	
}


\section{Results and discussion}\label{sec:results3}
\subsection{Validation}
\subsubsection{Grid resolution study and drag uncertainty estimation}
Grid convergence studies on the shock and interface capturing behavior of the scheme were performed by Garrick et al.~\cite{Garrick2016,Garrick2016a}.
In the present work, the effect of grid resolution on the breakup behavior and drag coefficient is examined with several simulations of the $M_s=3$, $\mathrm{We_a}=100$ (crossflow $\mathrm{We}\approx2300$) shock-column interaction with grid resolutions in the vicinity of the column of $D/60$, $D/120$, $D/240$, and $D/480$.
In lieu of performing an exhaustive grid resolution study at each shock speed and Weber number to be tested, it is assumed that relatively similar behavior trends will apply for the range of conditions in the production runs to follow.

\revb{First it is important to highlight the limitations of the present simulations. As noted by Jain et al.~\cite{Jain2015}, liquid breakup is ultimately a molecular process and without multiscale modeling the breakup will be initiated by the grid resolution. As noted by Meng and Colonius~\cite{Meng2018} in their recent paper, this means grid convergence of the breakup behavior is impossible to achieve in a traditional sense. With regards to the viscous effects, direct numerical simulations that resolve the boundary layer on the liquid surface are impractical without highly specialized solvers capable of both significant adaptive mesh refinement and additional body fitted structured conformal meshes which can achieve effective grid resolutions of up to $D/4000$~\cite{Chang2013}. For these reasons, recent studies of secondary atomisation in this flow regime have tended to consider flow conditions where viscous and surface tension effects can be safely neglected~\cite{Meng2018,Liu2018,Xiang2017}. Therefore while both viscous and surface tension effects are included in the simulations presented here, it should be acknowledged that these effects will be under-resolved to some degree.  However, the goal is partly to determine to what degree the physics involved in secondary atomisation can be captured despite this limitation. }

\begin{figure}\centering
	\subfigure[Drag coefficient over time for different grid resolutions.]{\includegraphics[width=0.45\textwidth]{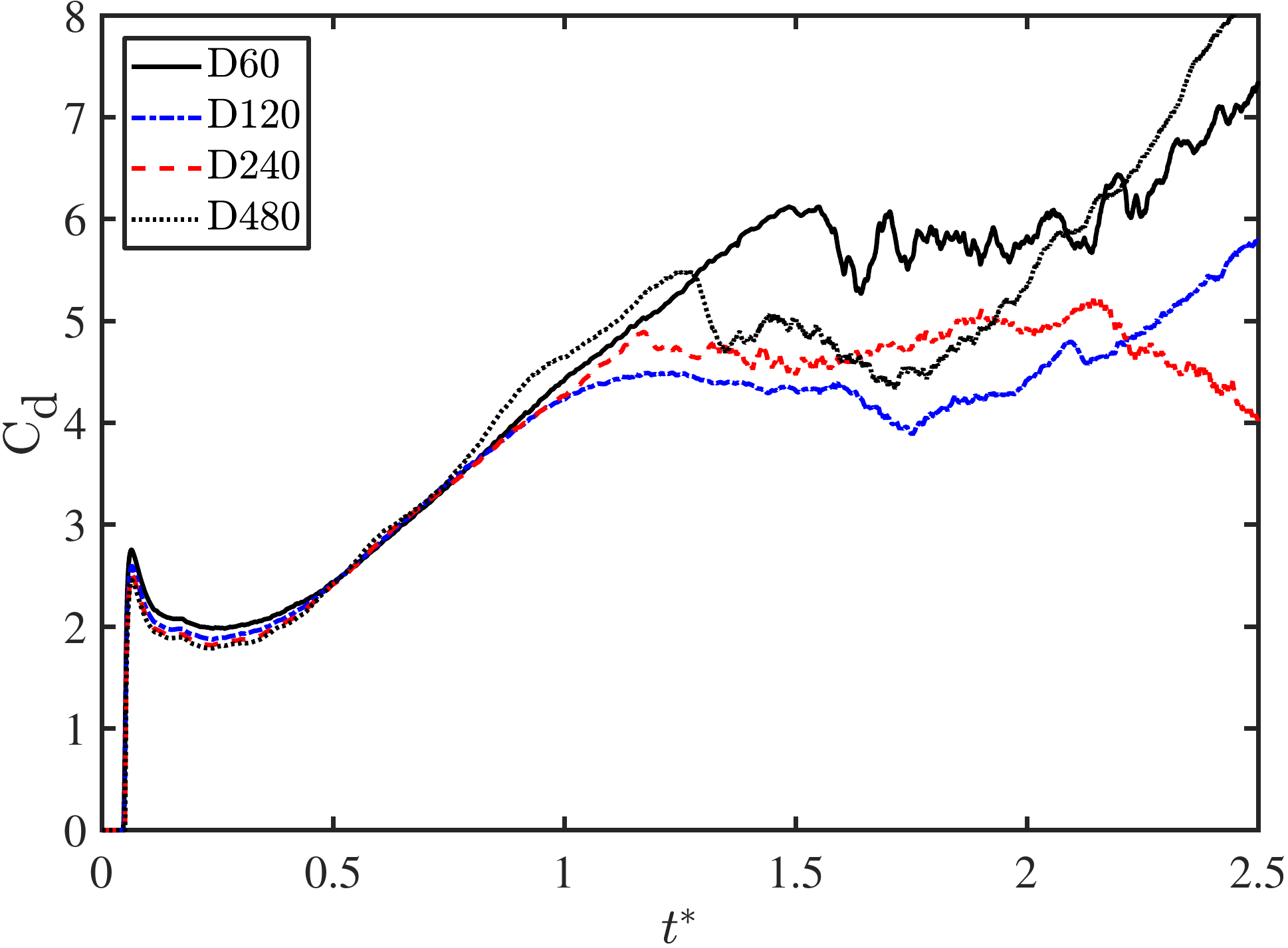}}
	\subfigure[\reva{Mean and standard deviation (SD) of drag coefficient from all grid resolutions at each time point as an estimate of drag coefficient uncertainty over time.}]{\includegraphics[width=0.45\textwidth]{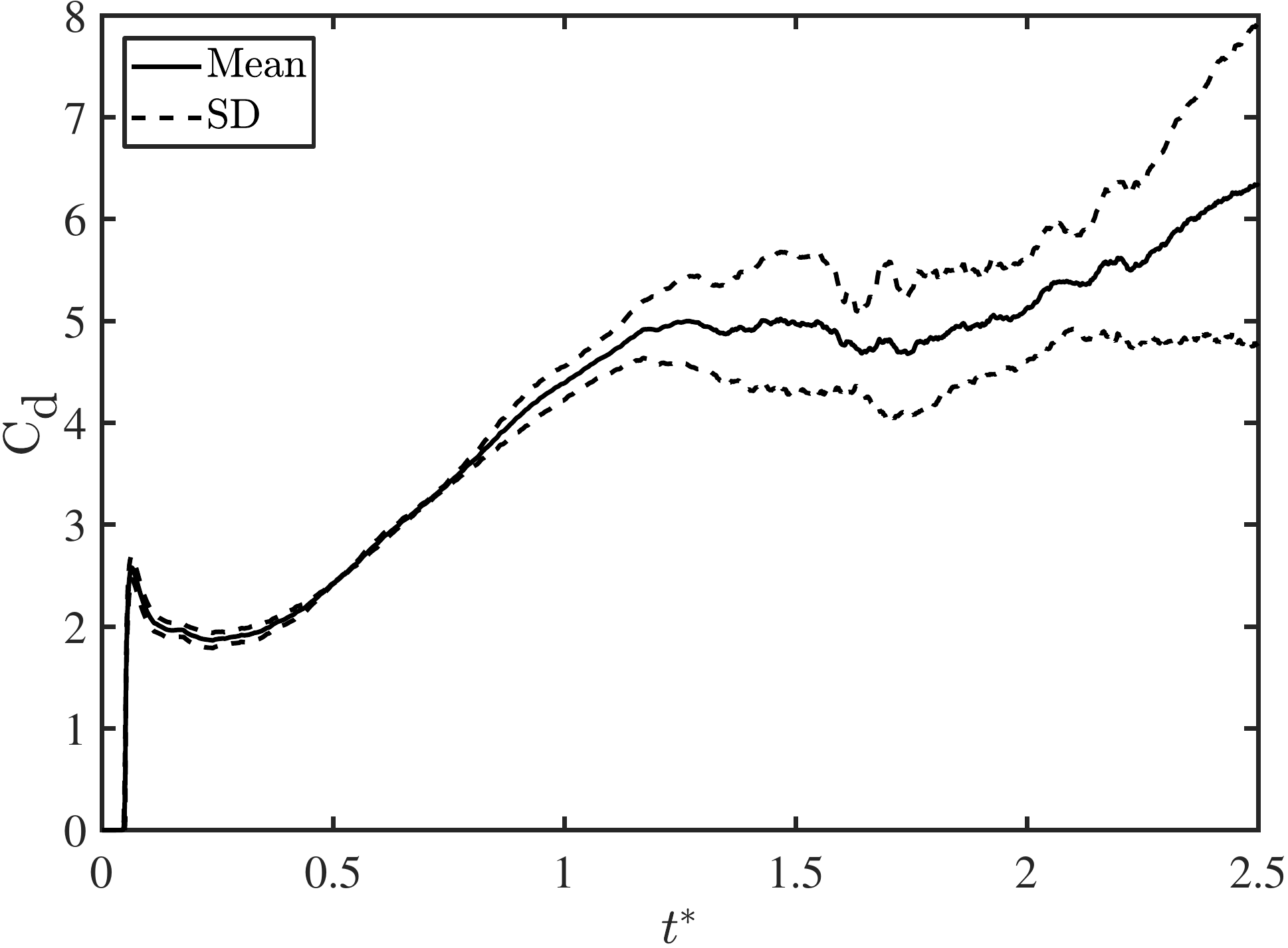}}
	\caption{\label{fig:cdgridreso}$M_s=3$, $\mathrm{We_a}=100$ drag coefficient at different grid resolutions (left) and an estimate of the uncertainty in the drag coefficient over time (right).}
\end{figure}

First, the drag coefficient is examined in Figure~\ref{fig:cdgridreso}.
Note that the drag (Eq.~\ref{eqn:cd}) is determined by integrating the acceleration of the total liquid mass in the domain (Eq.~\ref{eqn:ac}), so as liquid mass is separated and swept downstream it will have a corresponding effect on the drag coefficient.
This is particularly noticable in Figure~\ref{fig:cdgridreso} where the drag coefficients separate around $t^*=1$, however, they remain reasonably correlated until approximately $t^*=2$ at which point they diverge.

\begin{figure}\centering
	\includegraphics[width=0.85\textwidth,right]{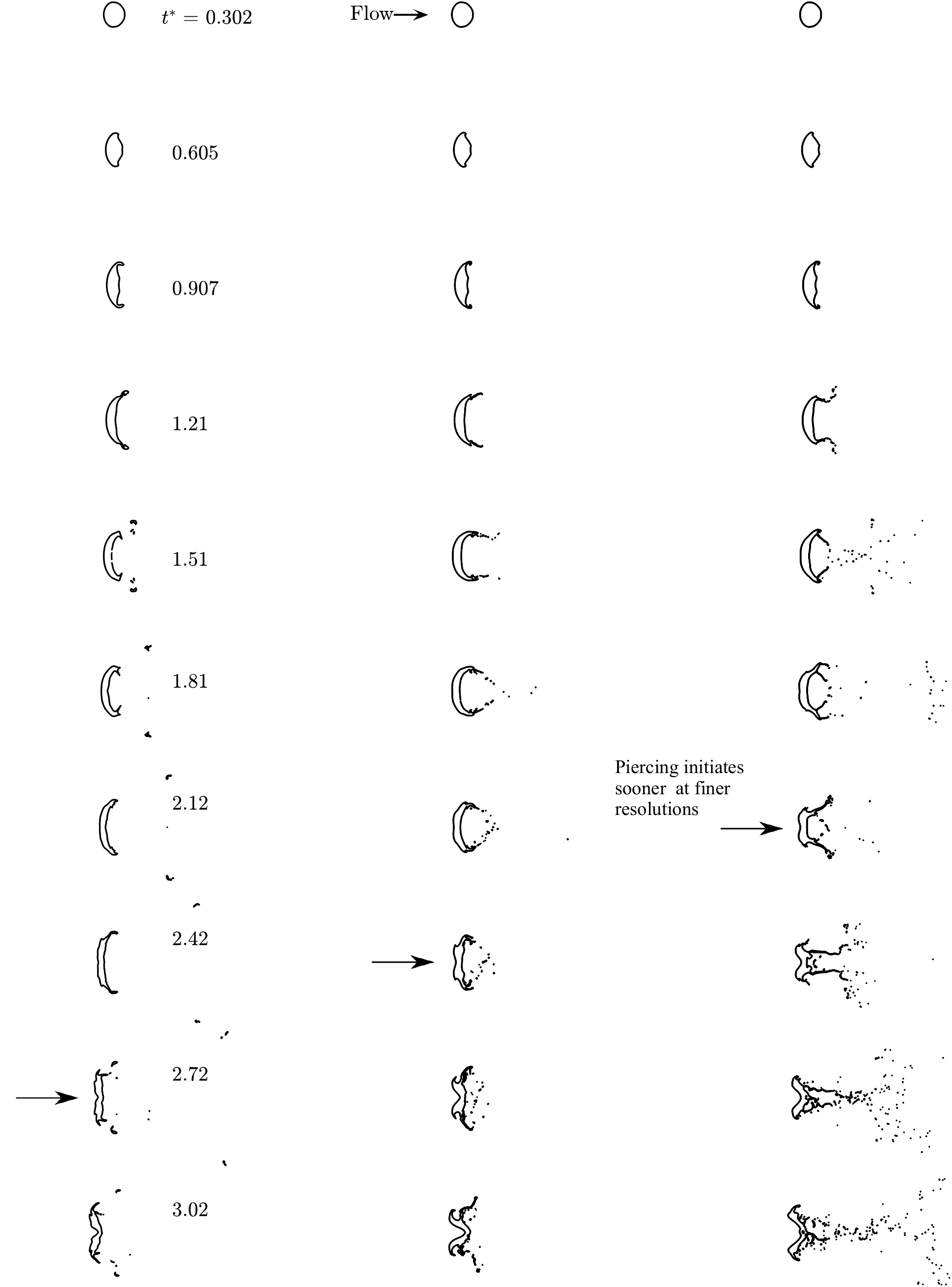}
	\caption{\label{fig:gridresoMs3}$M_s=3$, $\mathrm{We_a}=100$ breakup behavior at $D/60$ (left), $D/120$ (center), and $D/240$ (right) grid resolutions.}
\end{figure}

 The deformation and breakup behavior of the different simulations is depicted in Figure~\ref{fig:gridresoMs3} which depicts a time history of the gas-liquid interface (i.e.
$\phi_1=0.5$ iso-line) throughout the simulations where each row depicts a different solution time and each column a different grid resolution.
Like with the drag coefficient, the early stages ($t^*<1$) of the deformation process does not vary significantly across the grid resolutions tested.
For $1< t^*<2$ more fine scale ligament and droplet features are observed in the finer grid resolutions but the general behavior remains similar in the three simulations.
The minor differences in the location and trajectory of the smaller droplet particles impact the computed drag coefficient and explains the previously discussed separation of the coefficients in Figure~\ref{fig:cdgridreso} for $t^*>1$.
 For $t^*>2$ the general behavior consists of the flow ``piercing" through the center of the droplet.
This piercing is initiated sooner at the finer grid resolutions (or delayed on coarser grids) but the general breakup behavior is qualitatively similar in all three simulations, albeit with significantly more small droplets captured on the finest grid. 

\revb{Finally, an additional $D/120$ simulation was performed with a domain twice as large and produced nearly identical results to the original $D/120$ simulation, verifying the domain size was not impacting the results.}

These results can be broken into several useful groups based on the observed behavior of the drag coefficient and breakup characteristics.
For $t^*<1$ the results converge and should provide a reasonable estimate of the drag coefficient and droplet deformation.
From $1\leq t^*< 2$ there is some uncertainty in the breakup behavior in terms of the presence and trajectory of smaller droplet clouds, however the general behavior remains the same and as the drag coefficients reasonably correlate across the grid resolutions they should provide at least a first order estimate.
For $t^*\geq 2$ there is significantly more uncertainty in the drag coefficients which begin to diverge across the grid resolutions, however, the general breakup behavior is still observed at all three resolutions.


\subsection{Deformation and breakup behavior}\label{sec:breakup}
The effect of Weber number on the deformation and breakup characteristics of the liquid column is investigated for each shock speed \revb{using a grid resolution of D/120.}
Time histories of the gas-liquid interface (i.e.
$\phi_1=0.5$ iso-line) are shown in corresponding figures where the Weber numbers are depicted at the bottom of each figure.
Each row depicts a different solution time and each column a different Weber number.
The characteristic time $t^*$ for each row of images is depicted on the left side of each figure.
In all cases the crossflow is traveling from left to right.


\subsubsection{$M_s=1.47$}
\begin{figure}\centering
	\includegraphics[width=0.95\textwidth,right]{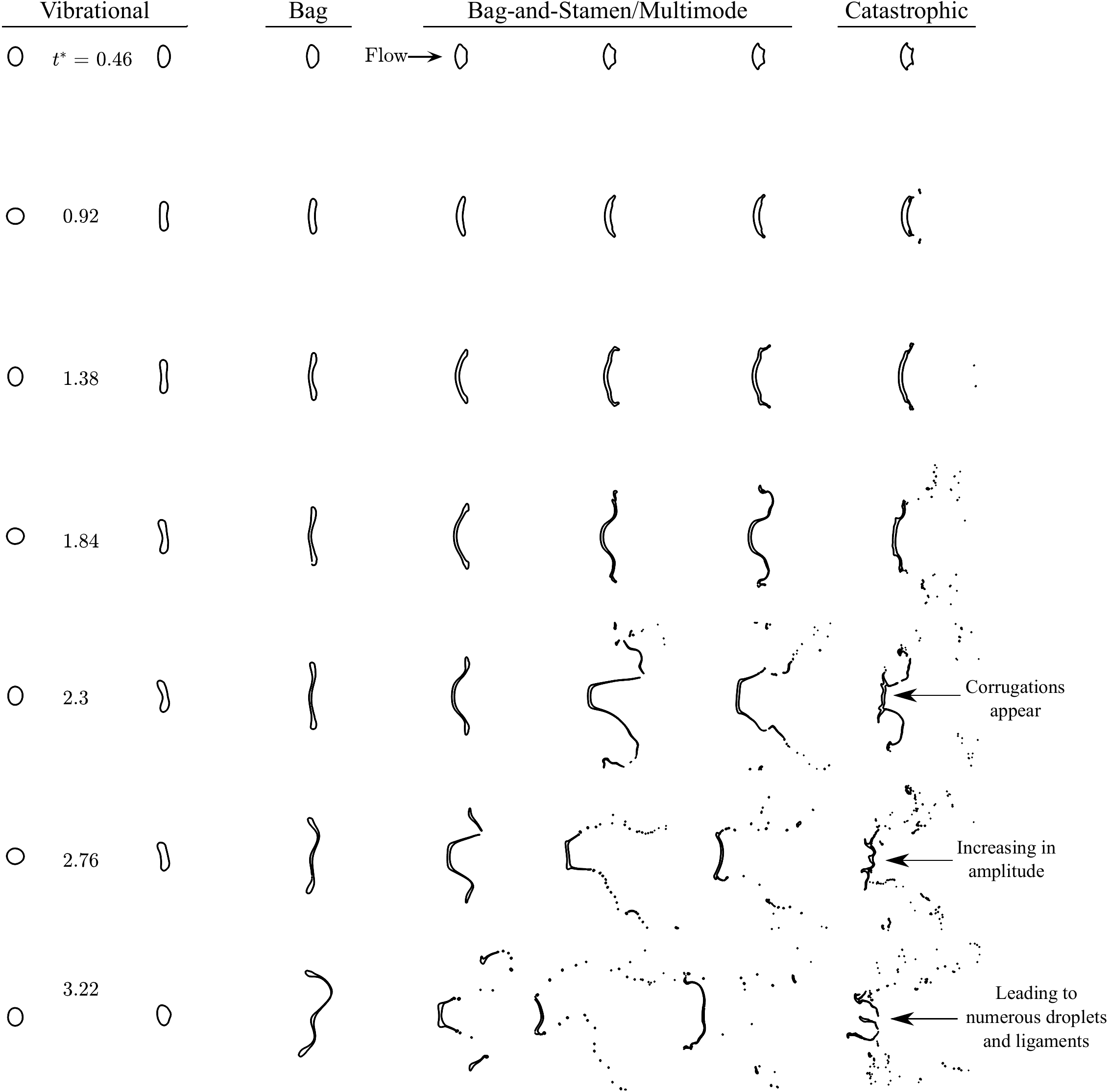} 
	\begin{tabular*}{\textwidth}{ll @{\extracolsep{\fill}} rrrrrrrr}
	  & (a)     & (b)      & (c)   & (d)& (e)& (f)& (g) & \quad \\
		$\mathrm{We_a}$ & 1     & 5      & 10   & 20& 50& 100& 1000 & \quad \\
		$\mathrm{We_c}$ & 0.9 & 4.7 & 9.4 & 19& 47& 94& 941 & \quad \\
	\end{tabular*}
	\caption{\label{fig:Ms147}$M_s=1.47$ deformation and breakup behavior.}
\end{figure}

Figure~\ref{fig:Ms147} depicts the results for the $M_s=1.47$ simulations.
For this Mach number, the crossflow Weber numbers correspond closely to the acoustic Weber numbers.\revb{ For this shock strength the local gas-liquid density ratio using the initial post-shock gas conditions is $\rho_l/\rho_g\approx460$.}
The observed breakup characteristics exhibit reasonable qualitative agreement with the different regimes observed in subsonic experiments for $\mathrm{Oh<0.1}$.
The regimes are listed in Table~\ref{tab:regimes} where the transition Weber numbers are approximate partly due to the continuous nature of the breakup process and the arbitrary choice for specific transition points~\cite{Guildenbecher2009}.
As a result, different researchers have reported slight variations on the transition between different regimes~\cite{Pilch1987}, however, the order in which they appear remains the same~\cite{Jain2015}.
 At lower Weber numbers (Figure~\ref{fig:Ms147}(a) and (b)) a vibrational type mode is observed where the \reva{surface tension is} large enough for the column to remain intact and oscillate as an ellipse.

\begin{table}\centering
	\caption{Breakup regimes and transition Weber number as given by~\cite{Guildenbecher2009}.\label{tab:regimes}}
	\begin{tabular}{cr}
		\hline
		Vibrational & $0 < \mathrm{We} < \sim 11$ \\
		Bag & $\sim 11 < \mathrm{We} < \sim 35$ \\
		Multimode & $\sim 35 < \mathrm{We} < \sim 80$ \\
		Sheet-thinning & $\sim 80 < \mathrm{We} < \sim 350$ \\
		Catastrophic &  $\mathrm{We} > \sim 350$
	\end{tabular}
\end{table}

 Figure~\ref{fig:Ms147}(c) depicts various stages of what appears to be a bag breakup process.
Generally this regime is characterized by the growth of a bag structure where the center of the drop is blown downstream and attached to an outer rim.

In the bag-and-stamen/multi-mode regime, the center of the droplet is driven downstream more slowly than the rim leading to the creation of a bag/plume structure~\cite{Dai2001}.
Similar features are observed in the present liquid column simulations as depicted in Figure~\ref{fig:Ms147}(d) and (e).  Figure~\ref{fig:Ms147}(d) depicts the formation of this bag-and-stamen type structure at a slightly lower Weber number (20) compared to the breakup regimes observed for incompressible flow characterized in Table~\ref{tab:regimes}. However in the present compressible flow simulations, a small standing shock is observed downstream of the liquid column. A similar standing shock feature has been observed in prior numerical results without surface tension at this flow speed~\cite{Meng2014,Terashima2009}.
The pressure disturbance caused by the presence of the standing shocks could contribute to the growth of the bag-and-stamen structure. Figure~\ref{fig:Ms147}(e) is characterized by a substantial plume/bag-and-stamen structure forming around $t^*=2.3$ before its subsequent rupture into numerous small droplets.
Finally, the breakup characteristics in Figure~\ref{fig:Ms147}(g) correlate well with the so-called catastrophic regime where the drop surface is corrugated by large amplitude waves resulting in a large number of smaller droplets and ligaments~\cite{Guildenbecher2009}.



\subsubsection{$M_s=2$}

\begin{figure}\centering
	\includegraphics[width=0.95\textwidth,right]{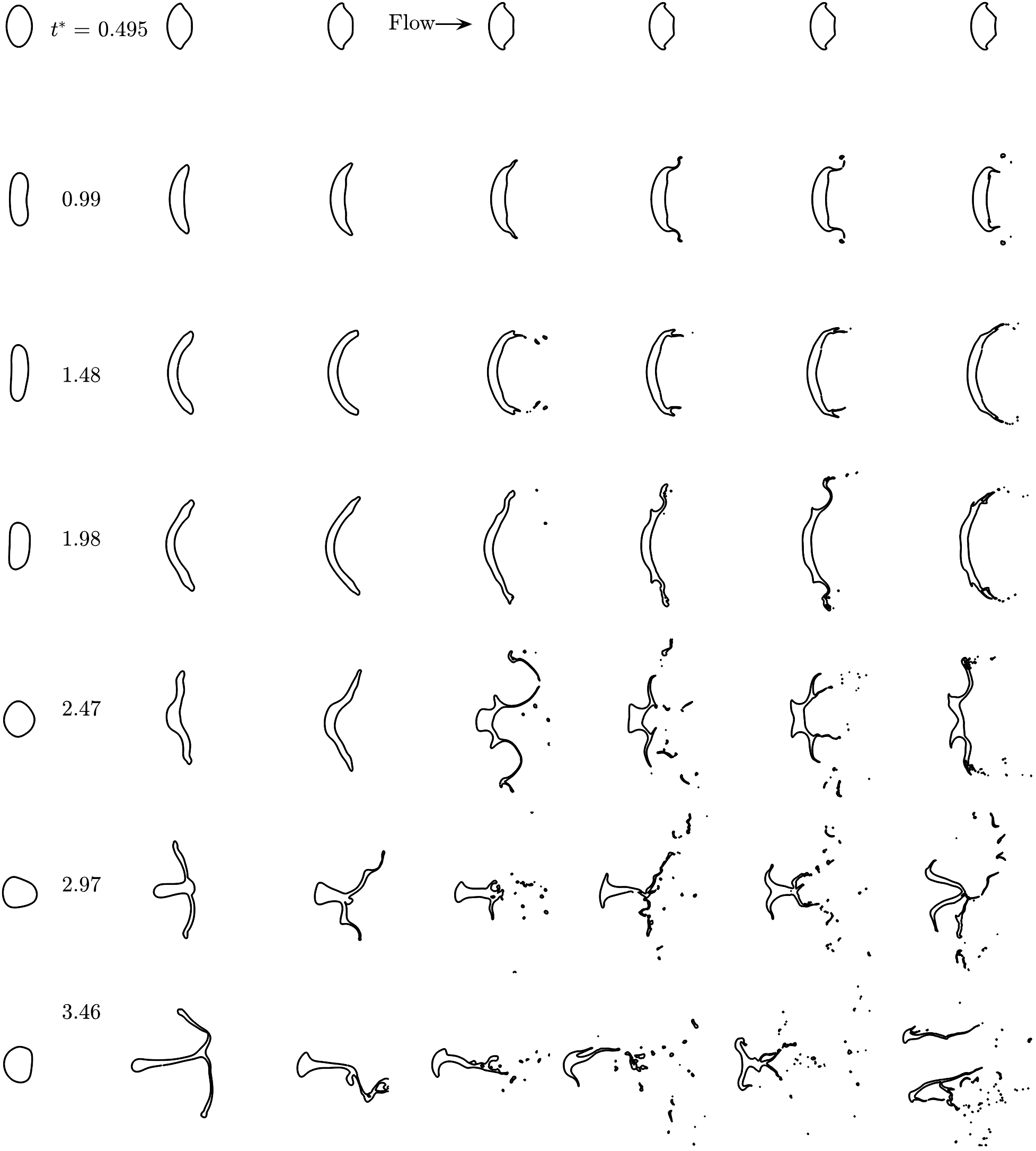} 
	\begin{tabular*}{\textwidth}{cc @{\extracolsep{\fill}} rrrrrrrr}
			  & (a)   \hspace*{5mm}  & (b)   & (c)   & (d)& (e)& (f)& (g) \\
		$\mathrm{We_a}$ & 1 \hspace*{5mm} & 5      & 10   & 20& 50& 100& 1000 \\
		$\mathrm{We_c}$ & 5 \hspace*{5mm}& 25 & 50 & 100& 250& 500& 5000 \\
	\end{tabular*}
	\caption{\label{fig:Ms2}$M_s=2$ deformation and breakup behavior.}
\end{figure}

Figure~\ref{fig:Ms2} depicts the breakup behavior for the $M_s=2$ simulations.
The post-shock conditions are in the transonic regime with a crossflow Mach number of $M=0.96$. \revb{ For this shock strength the local gas-liquid density ratio using the initial post-shock gas conditions is $\rho_l/\rho_g\approx312$.}
Across the range of Weber numbers the breakup behavior is very similar to the slower $M_s=1.47$ case even as late as $t^*=2$.
However, at later times the general breakup behavior begins to noticably deviate from the lower Mach number case, especially with respect to the overall size of the ligament structures which were observed to stretch considerably further in the $M_s=1.47$ simulations.
As the $M_s=2$ shock induces a faster crossflow than the $M_s=1.47$ case, the crossflow Weber number corresponding to each acoustic Weber number is slightly higher.
Figure~\ref{fig:Ms2}(b) depicts a bag-and-stamen type breakup structure with the outer rim of the column being swept downstream faster than the center of the column, resulting in the formation of several ligament structures.
Figures~\ref{fig:Ms2}(c)-(e) depict a unique multimode type of asymmetric breakup culminating in the collapse of the droplet into a largely coherent ligament structure although an increasing number of smaller droplets are generated during this process at the higher Weber numbers.
This noticably asymmetric behavior appears to originate from small asymmetries which appear earlier during the deformation process, i.e.
in Figures~\ref{fig:Ms2}(c)-(e) at $t^*=1.98, 2.47$.
 Finally, a catastrophic type breakup is observed at the highest Weber numbers in Figures~\ref{fig:Ms2}(f) and (g).

\subsubsection{$M_s=2.5$}

\begin{figure}\centering
	\includegraphics[width=0.95\textwidth,right]{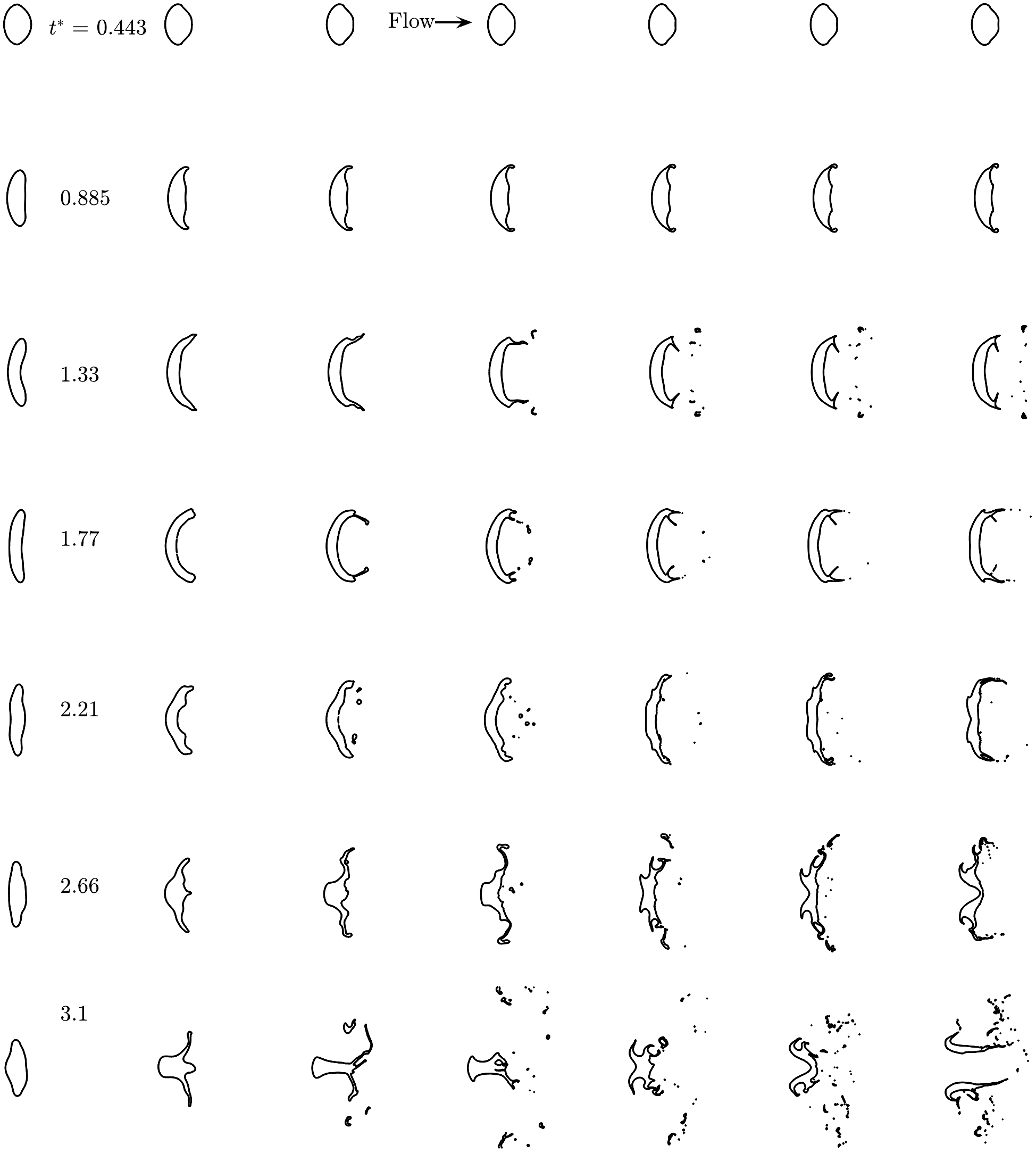} 
	\begin{tabular*}{\textwidth}{cc @{\extracolsep{\fill}} rrrrrrr}
		\hspace*{3mm}	  & (a)  \hspace*{3mm}   & (b) \hspace*{3mm}     & (c) \hspace*{3mm}  & (d) \hspace*{3mm}& (e) \hspace*{3mm}& (f) \hspace*{3mm}& (g) \\
	\hspace*{3mm}	$\mathrm{We_a}$ & 1  \hspace*{3mm}   & 5   \hspace*{3mm}   & 10 \hspace*{3mm}  & 20 \hspace*{3mm}& 50 \hspace*{3mm}& 100 \hspace*{3mm}& 1000 \\
	\hspace*{3mm}	$\mathrm{We_c}$ & 12\hspace*{3mm} & 61   \hspace*{3mm}   & 123 \hspace*{3mm} & 245 \hspace*{3mm}& 613 \hspace*{3mm}& 1225 \hspace*{3mm}& 12250 \\
	\hspace*{3mm}	$\mathrm{We_{eff}}$ & 9.2 \hspace*{3mm}& 46 \hspace*{3mm} & 92 \hspace*{3mm} & 183 \hspace*{3mm}& 458 \hspace*{3mm}& 917  \hspace*{3mm}  & 9167 \\
	\end{tabular*}
	\caption{\label{fig:Ms250}$M_s=2.5$ deformation and breakup behavior.}
\end{figure}

Figure~\ref{fig:Ms250} depicts the breakup behavior for the $M_s=2.5$ simulations.
 The higher incident shock speed means the post-shock conditions consist of a supersonic flow.\revb{ For this shock strength the local gas-liquid density ratio using the initial post-shock gas conditions is $\rho_l/\rho_g\approx250$.}
As a result, the estimated crossflow Weber number is much higher for each acoustic Weber compared to the corresponding $M_s=1.47$ and $M_s=2$ simulations.
With the presence of supersonic flow and an associated bow shock appearing in front of the droplet, the effective post-shock Weber number is computed using Eq.~\ref{eqn:weeff} to provide a comparable metric for subsonic simulations. \revb{Using the same approach to compute an effective gas-liquid density ratio accounting for the bow-shock gives $\rho_l/\rho_g\approx187$.}
The breakup behavior is generally similar to the $M_s=2$ simulations with a vibrational type mode observed in Figure~\ref{fig:Ms250}(a), multimode type behavior in Figures~\ref{fig:Ms250}(b)-(d) and catastrophic type breakup in Figures~\ref{fig:Ms250}(e)-(g).
Similarly to the $M_s=2$ simulations, a feature of this catastrophic breakup behavior is the generation of a ``channel" whereby the liquid column is pierced in the center into two separate chunks.

\subsubsection{$M_s=3$}

\begin{figure}
	\includegraphics[width=.95\textwidth,right]{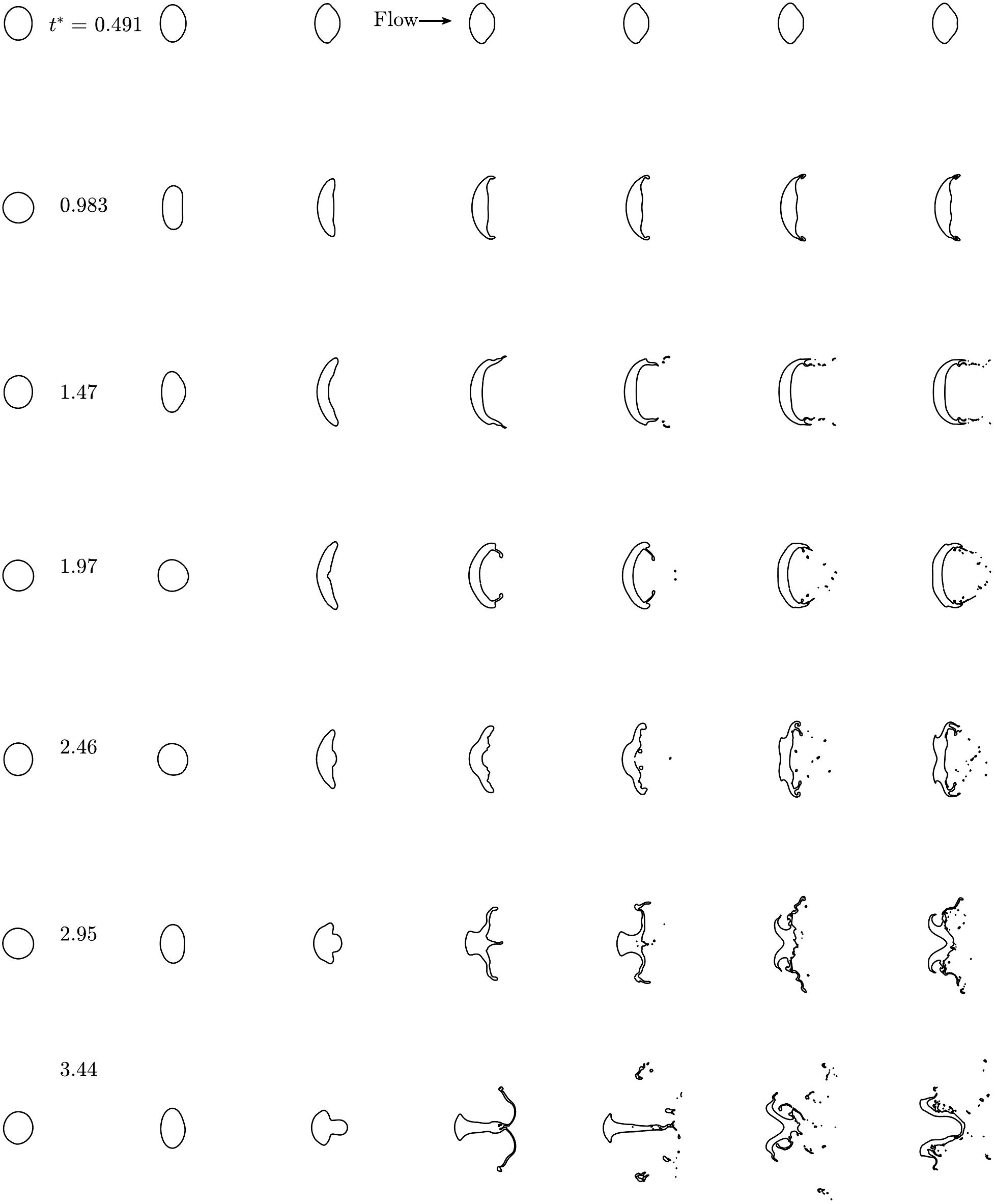} 
\begin{tabular*}{\textwidth}{cc @{\extracolsep{\fill}} rrrrrrr}
		  & (a)   \hspace*{5mm}  & (b)   \hspace*{5mm}   & (c) \hspace*{5mm}  & (d) \hspace*{5mm}& (e) \hspace*{5mm}& (f) \hspace*{5mm}& (g) \\
		$\mathrm{We_a}$ & 0.05 \hspace*{5mm} & 0.2  \hspace*{5mm}  &  1 \hspace*{5mm}  &  5 \hspace*{5mm}& 10 \hspace*{5mm}&  50 \hspace*{5mm}& 100  \\
		$\mathrm{We_c}$ & 1.1 \hspace*{5mm} & 4.6 \hspace*{5mm} &22.9 \hspace*{5mm}& 114 \hspace*{5mm}& 229 \hspace*{5mm}& 1143 \hspace*{5mm}&  2286 \\
		$\mathrm{We_{eff}}$ & 0.71 \hspace*{5mm} &2.8 \hspace*{5mm}&14.1 \hspace*{5mm}&  71 \hspace*{5mm}& 141 \hspace*{5mm}&   707 \hspace*{5mm}     &  1414 \\
	\end{tabular*}
	\caption{\label{fig:Ms3}$M_s=3$ deformation and breakup behavior.}
\end{figure}

Theofanous et al.~\cite{Theofanous2004} performed experiments of aerobreakup of spherical liquid droplets in $M=3$ crossflows.
They observed ``piercing" ($44<\mathrm{We}<10^3$) and ``stripping" ($\sim 10^3<\mathrm{We}$) breakup regimes.
Figure~\ref{fig:Ms3} depicts the breakup behavior for the present simulations which considers an $M_s=3$ shock speed that results in a considerably slower $M=1.36$ crossflow compared to the experiments of Theofanous et al.
Despite this difference, the range of breakup features depicted in Figure~\ref{fig:Ms3} with the estimated effective Weber numbers varying from approximately 0.7 in Figure~\ref{fig:Ms3}(a) to 1400 in Figure~\ref{fig:Ms3}(g)  appear to qualitatively match descriptions of the experimentally observed breakup regimes despite the disparity in crossflow speeds and flow dimensionality.
As with the previous simulations, the higher crossflow speed in the $M_s=3$ case results in significantly higher crossflow Weber numbers for each acoustic Weber number.
As a result, a significant number of small droplets are generated even at relatively low acoustic Weber numbers such as Figure~\ref{fig:Ms3}(e) and in the early stages of Figures~\ref{fig:Ms3}(f)-(g).
Catastrophic breakup is observed in the later stages of Figures~\ref{fig:Ms3}(f)-(g).
As in the $M_s=2$ and $M_s=2.5$ simulations, this catastrophic breakup is characterized by a channel which forms in the liquid column, splitting it into two.
This general behavior is similar to that experimentally observed for a waterdrop in a shocktube by Waldman et al~\cite{Waldman1972}. They described the breakup process as an initially continuous stripping of liquid from the droplet surface followed by a growth in the amplitude of surface waves which lead to the final disintegration of the droplet.
This description appears qualitatively similar to the time history of breakup depicted in Figures~\ref{fig:Ms3}(f)-(g).

\revb{ For this shock strength the local gas-liquid density ratio using the initial post-shock gas conditions is $\rho_l/\rho_g\approx216$. The effective gas-liquid density ratio accounting for the bow-shock gives $\rho_l/\rho_g\approx134$.}

\subsection{Drag coefficient}\label{sec:dragcoef}

\begin{figure}\centering
\subfigure[$M_s=1.47$]{\includegraphics[width=0.45\textwidth]{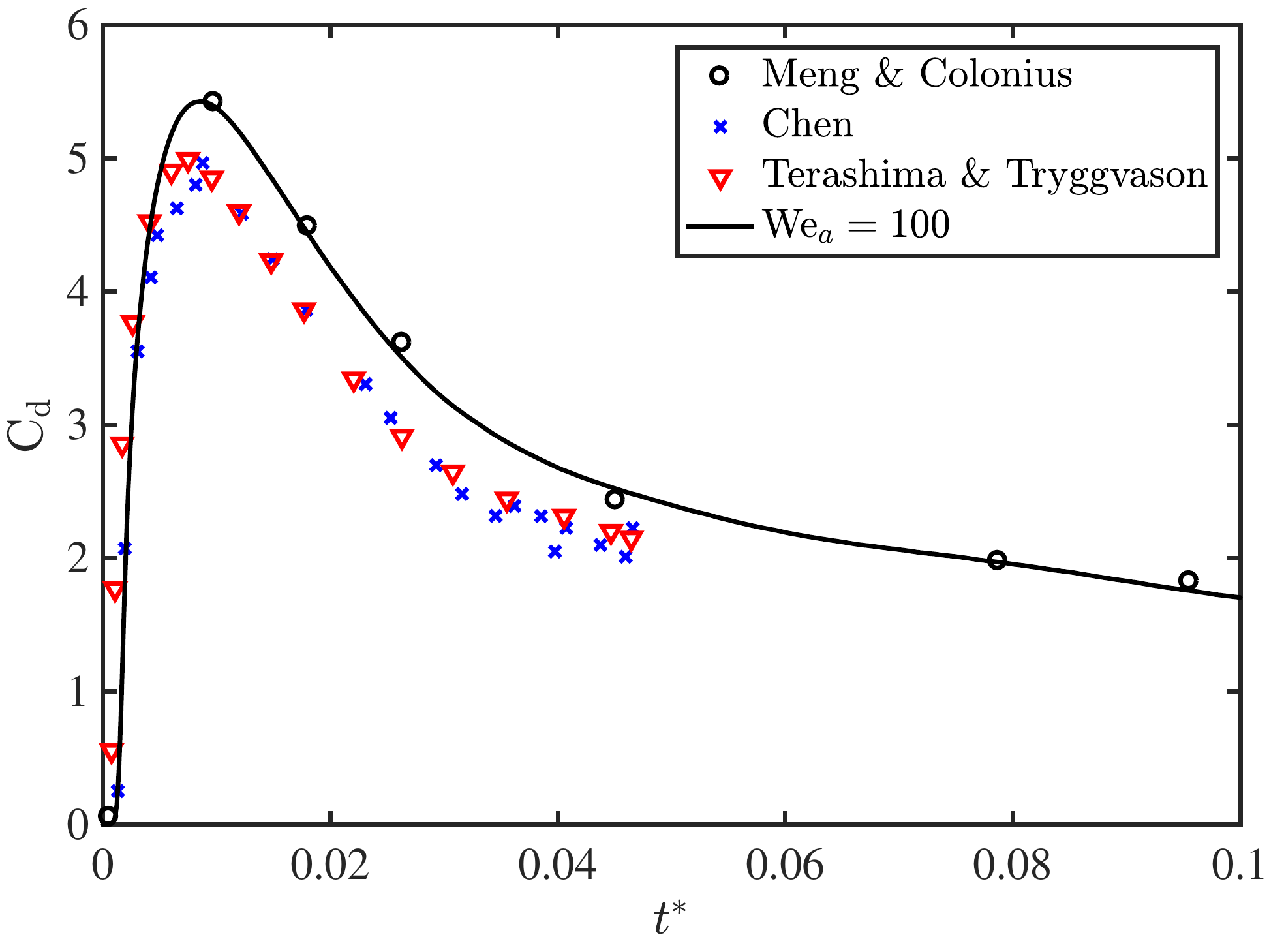}}
\subfigure[$M_s=2.5$]{\includegraphics[width=0.45\textwidth]{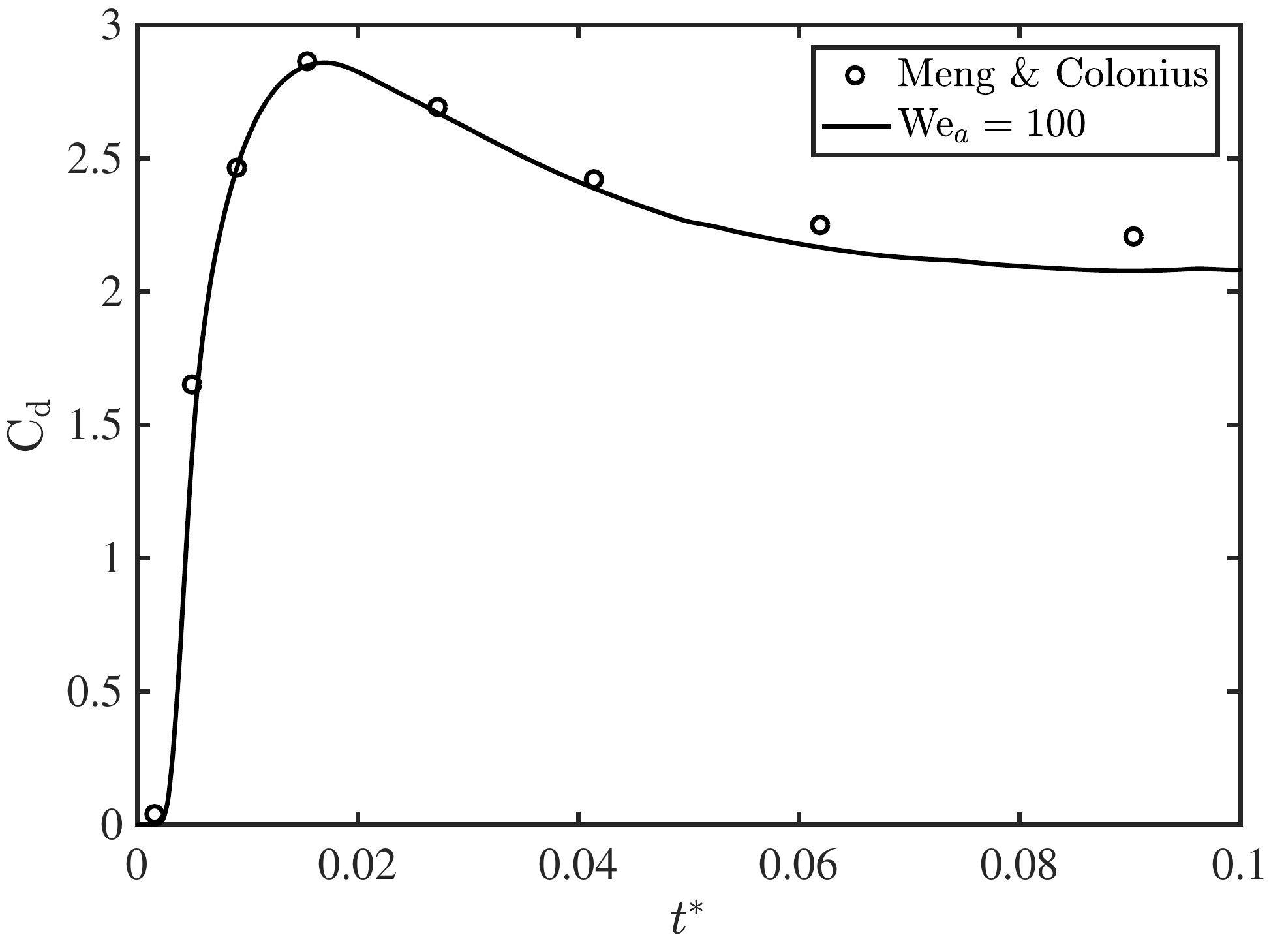}}
\caption{\label{fig:cde}Drag coefficient comparison during the early stages for $M_s=1.47$ (left) and $M_s=2.5$ (right) compared to Meng and Colonius~\cite{Meng2014}, Chen~\cite{Chen2008}, and Terashima and Tryggvason~\cite{Terashima2009}.}
\end{figure}

Figure~\ref{fig:cde} depicts comparisons of the early stages of the drag coefficient with prior numerical results of Meng and Colonius~\cite{Meng2014}, Chen~\cite{Chen2008}, and Terashima and Tryggvason~\cite{Terashima2009}.
The drag coefficient was computed following the approach of Meng and Colonius~\cite{Meng2014} as discussed in section~\ref{sec:drag}.
Good agreement is obtained with the data of~\cite{Meng2014}, disparities in the other results can likely be attributed to the use of a different approach to calculate the drag coefficient, where drift data (and not averaged fluid velocity) is used to estimate the column acceleration.
Further discussion of different approaches for computing the drag coefficient can be found in~\cite{Igra2002} and~\cite{Meng2014}.

\begin{figure}\centering
	\subfigure[$M_s=1.47$\label{fig:cd1}]{\includegraphics[width=0.45\textwidth]{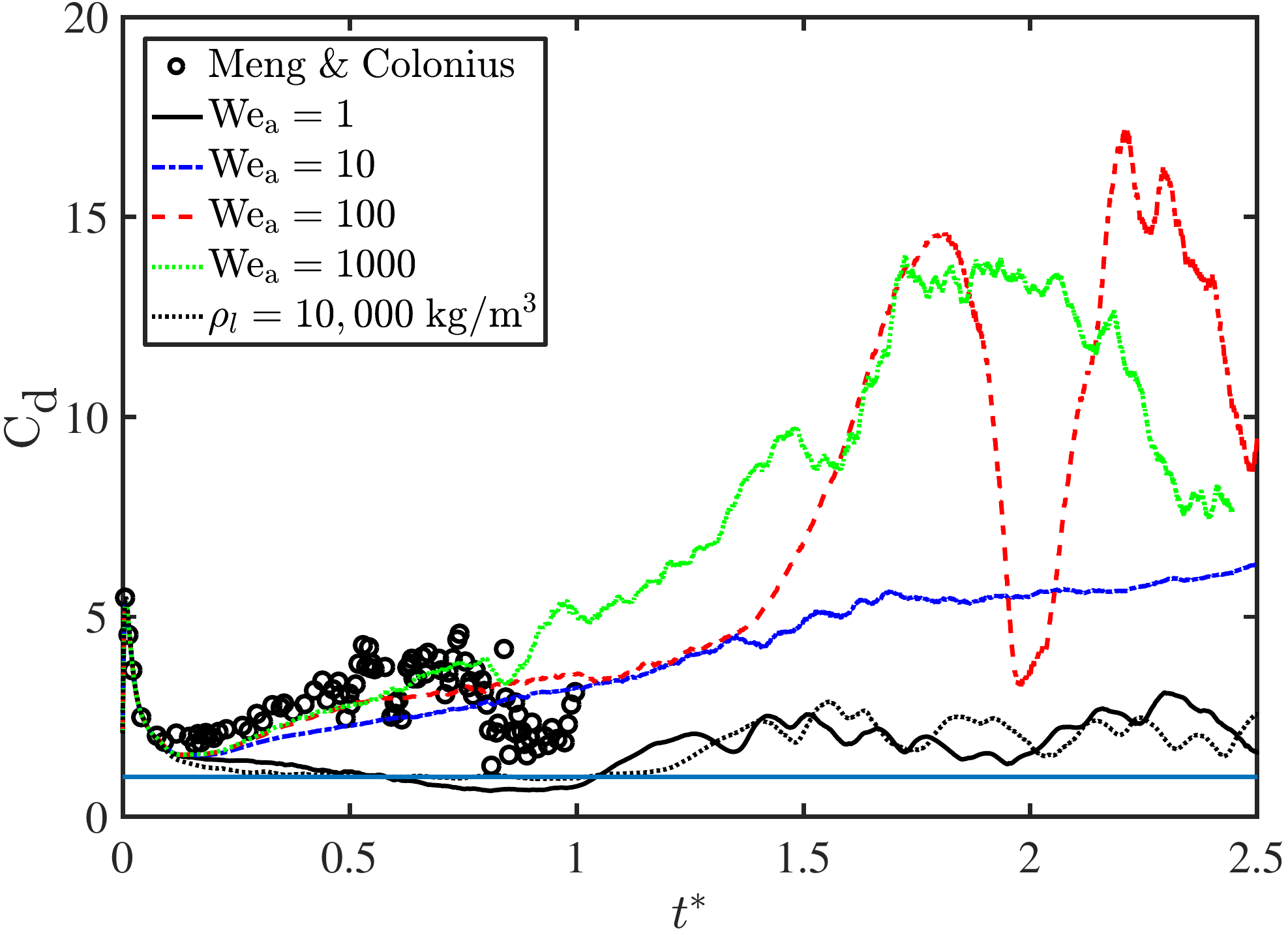}} 	\subfigure[$M_s=2$\label{fig:cd2}]{\includegraphics[width=0.45\textwidth]{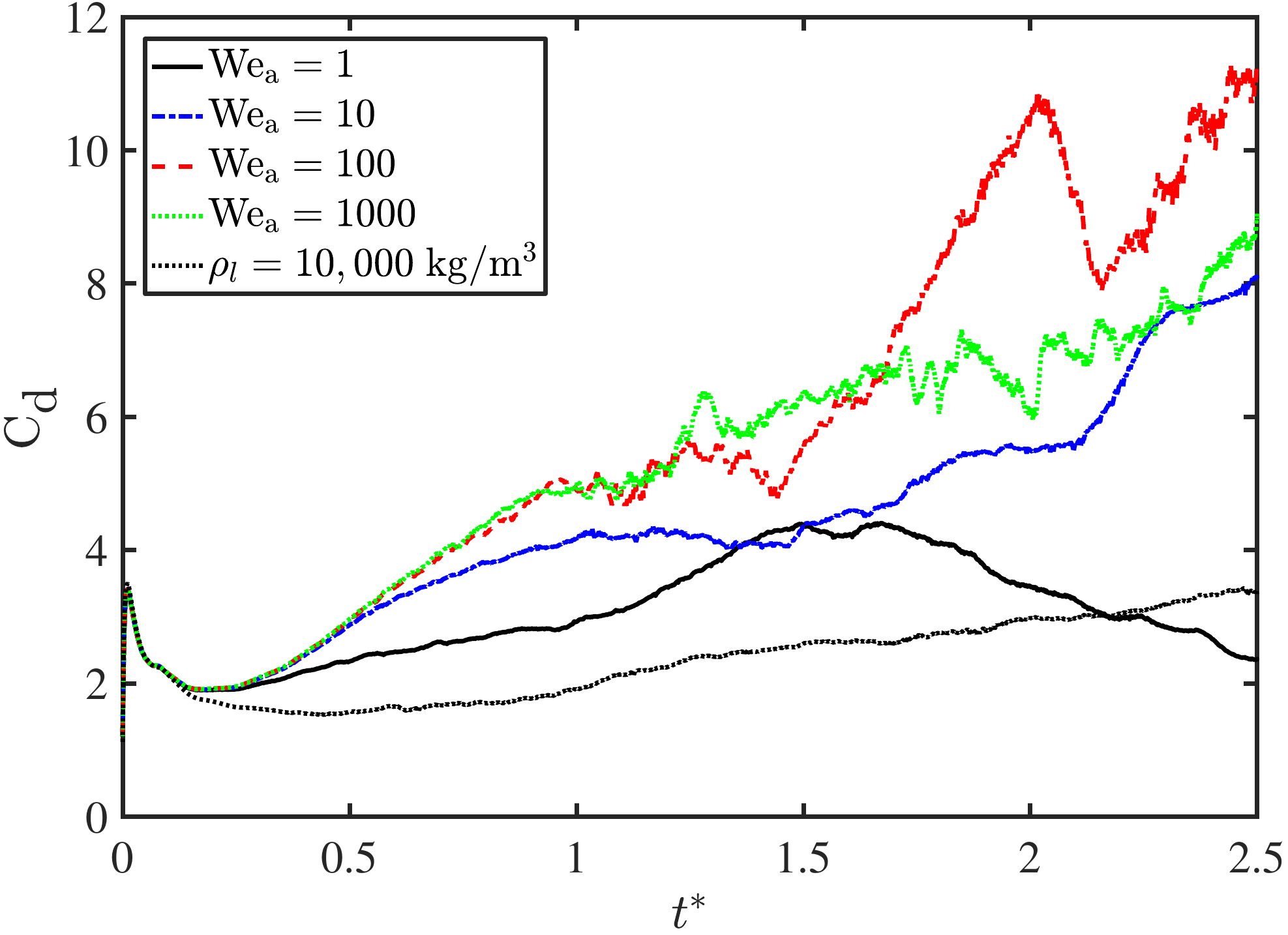}}\\
	\subfigure[$M_s=2.5$\label{fig:cd3}]{\includegraphics[width=0.45\textwidth]{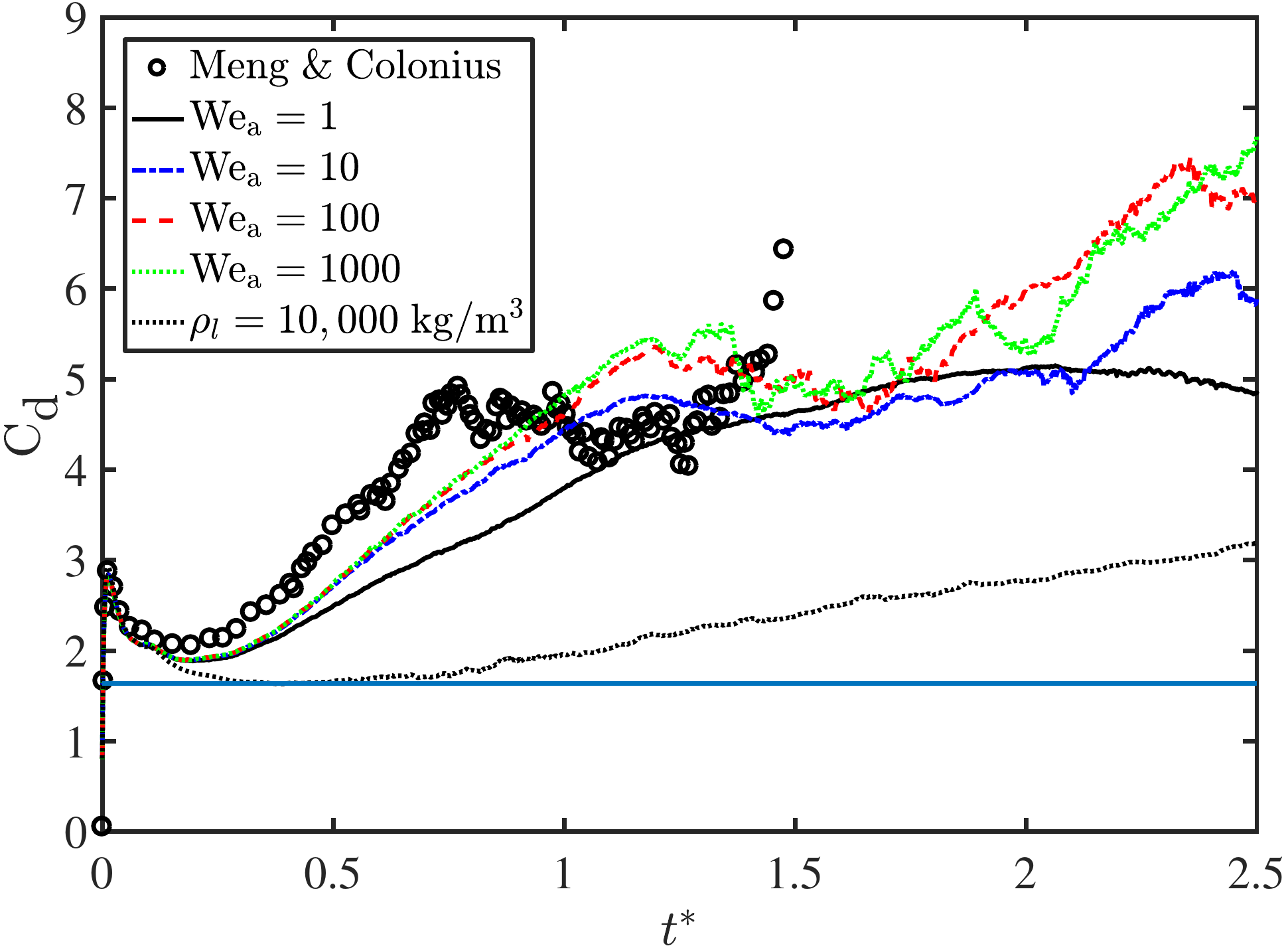}}
		\subfigure[$M_s=3$\label{fig:cd4}]{\includegraphics[width=0.45\textwidth]{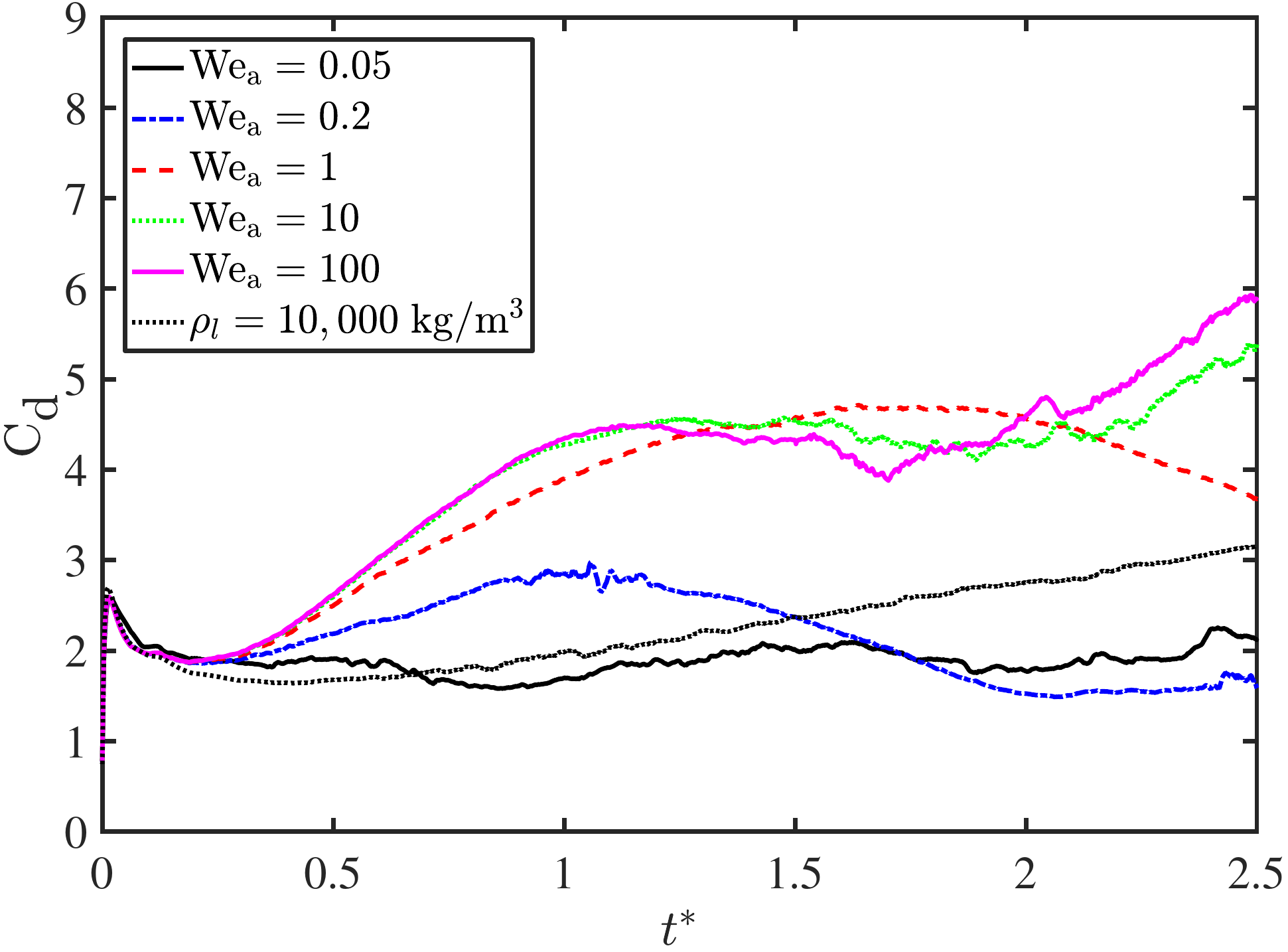}}
	\caption{\label{fig:cd}Drag coefficient comparison at the later stages.
The $M_s=1.47$ (a) and $M_s=2.5$ (c) cases include comparisons to Meng and Colonius~\cite{Meng2014}.
The solid blue line depicts approximate equivalent drag coefficients of a solid circular cylinder with $C_d\approx 1$ in (a) and $C_d\approx 1.64$ in (c).}
\end{figure}

Figure~\ref{fig:cd} depicts the drag coefficient at the later stages of the simulations with comparisons to Meng and Colonius~\cite{Meng2014}.
An extra simulation was also performed to provide a reference point to a stationary and rigid cylinder in crossflow where the drag coefficient is known.
This was approximated with a high liquid density ($\rho_l=10,000\mbox{ kg}/\mbox{m}^3$) case with $\mathrm{We_a}=1$.
Note that even under these conditions, some deformation of the high density liquid does occur.
Generally for $1000 < \mathrm{Re} < 3\times 10^5$, the drag coefficient of a cylinder is known to be approximately unity~\cite{Anderson}.
This value is plotted as a solid blue line in Figure~\ref{fig:cd1} and agrees well with the present subsonic simulation with a crossflow Reynolds number of 1430.
Meanwhile from Gowen and Perkins~\cite{Gowen1963} the drag coefficient of a stationary cylinder in a $M=1.2$ crossflow (i.e.
the crossflow for $M_s=2.5$) is approximately 1.64 and is plotted as a solid blue line in Figure~\ref{fig:cd3} for reference.
This value reasonably predicts the minimum drag coefficient value for the $\rho_l=10,000\mbox{ kg}/\mbox{m}^3$ simulation and which occurs around $t^*=0.3-0.4$ in Figure~\ref{fig:cd3}.

\revb{While the general trend is similar, overall the drag coefficient exhibits less unsteady variation compared to the results of~\cite{Meng2014}. The inclusion of surface tension and especially interface sharpening employed in the current simulations reduces the amount of liquid material stripped from the interface where it would otherwise enter the highly chaotic wake region and contribute to unsteady liquid acceleration measurements. }
Generally, lower drag coefficients are observed with lower Weber numbers for each shock Mach number except $M_s=3$ which shows less relative variation between the drag coefficients at Weber numbers in the range of 1 to 100 in Figure~\ref{fig:cd4}.

Significant differences in the drag as a function of the Weber number are observed in the $M_s=1.47$ and $M_s=2$ cases in Figures~\ref{fig:cd1} and~\ref{fig:cd2}.
Less variation is observed between the higher Weber numbers for the $M_s=2.5$ and $M_s=3$ cases depicted in Figure~\ref{fig:cd3} and~\ref{fig:cd4}.
Gowen and Perkins also noted there was almost no observed variation in the drag coefficient as a function of Reynolds number in the supersonic flow regime for a solid circular cylinder~\cite{Gowen1963}.
They stated that the suction pressures on the downstream side of the cylinder contribute a large part of the total drag in subsonic flows but as a percentage of the total drag this contribution rapidly decreases as the Mach number increases.

Supporting the experimental observations of Temkin and Mehta~\cite{temkin1982}, the unsteady drag is found to be larger in the decelerating relative flows of the liquid columns compared to that of the rigid stationary column.
The coefficients are observed to be twice as large or more compared to the rigid case for all shock Mach numbers.

Interestingly, comparing the present supersonic cases to the subsonic cases shows that at higher Mach numbers there is significantly less variation in the drag coefficient as a function of the Weber number for the liquid columns.
Upon first inspection, this is perhaps surprising as section~\ref{sec:breakup} demonstrated a broad range of breakup behaviors at each Mach number as a function of the Weber number and the drag is computed as an integration over the acceleration of the total liquid volume as it undergoes breakup.
However, an examination of the breakup behaviors for the supersonic cases in Figures~\ref{fig:Ms250} and \ref{fig:Ms3} appears to show a similar deformed diameter progression for the Weber number 1-100 cases within the respective Mach numbers.

\begin{figure}\centering
	\subfigure[$M_s=1.47$\label{fig:deff1}]{\includegraphics[width=0.45\textwidth]{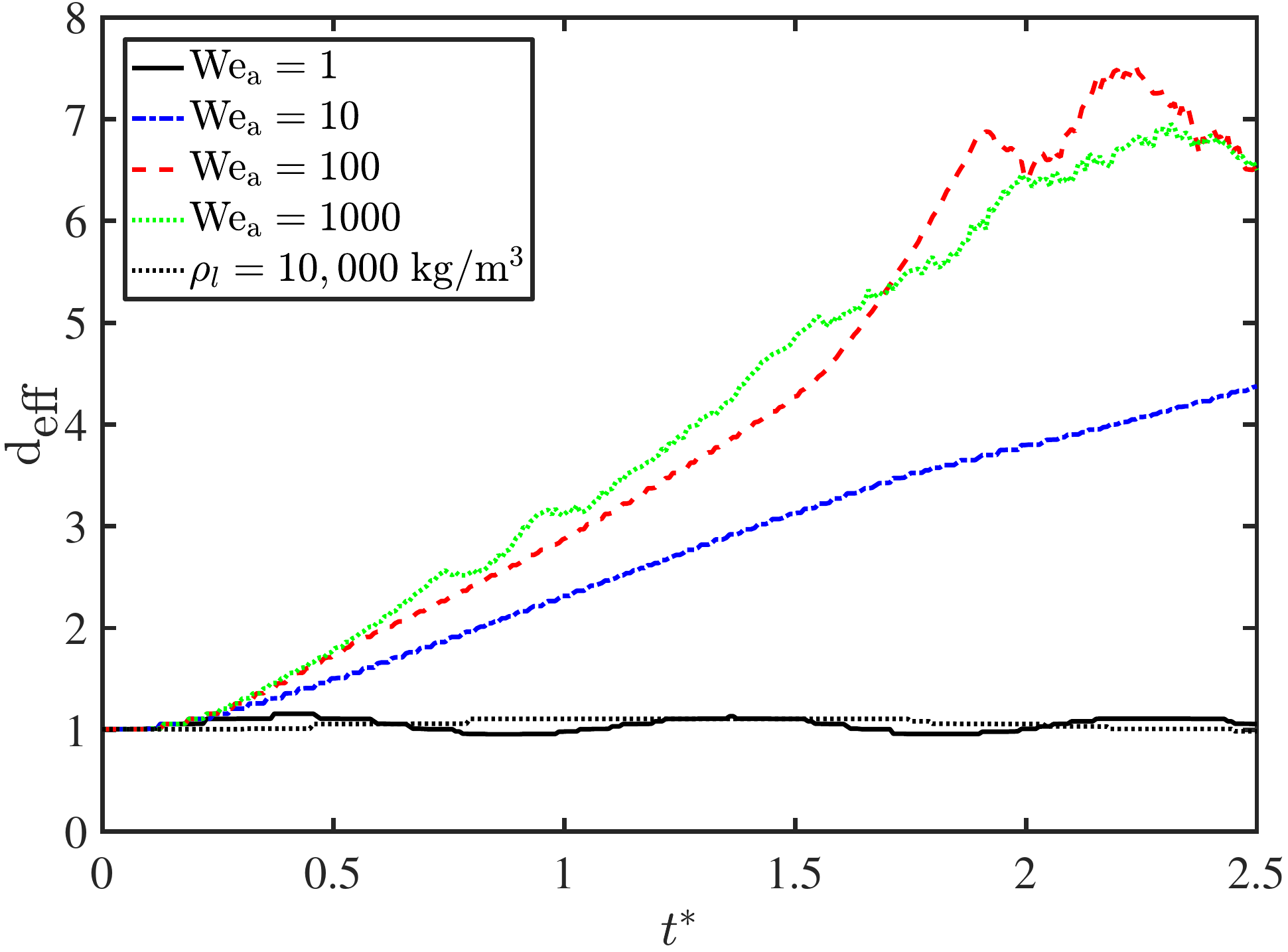}} 
	\subfigure[$M_s=2$\label{fig:deff2}]{\includegraphics[width=0.45\textwidth]{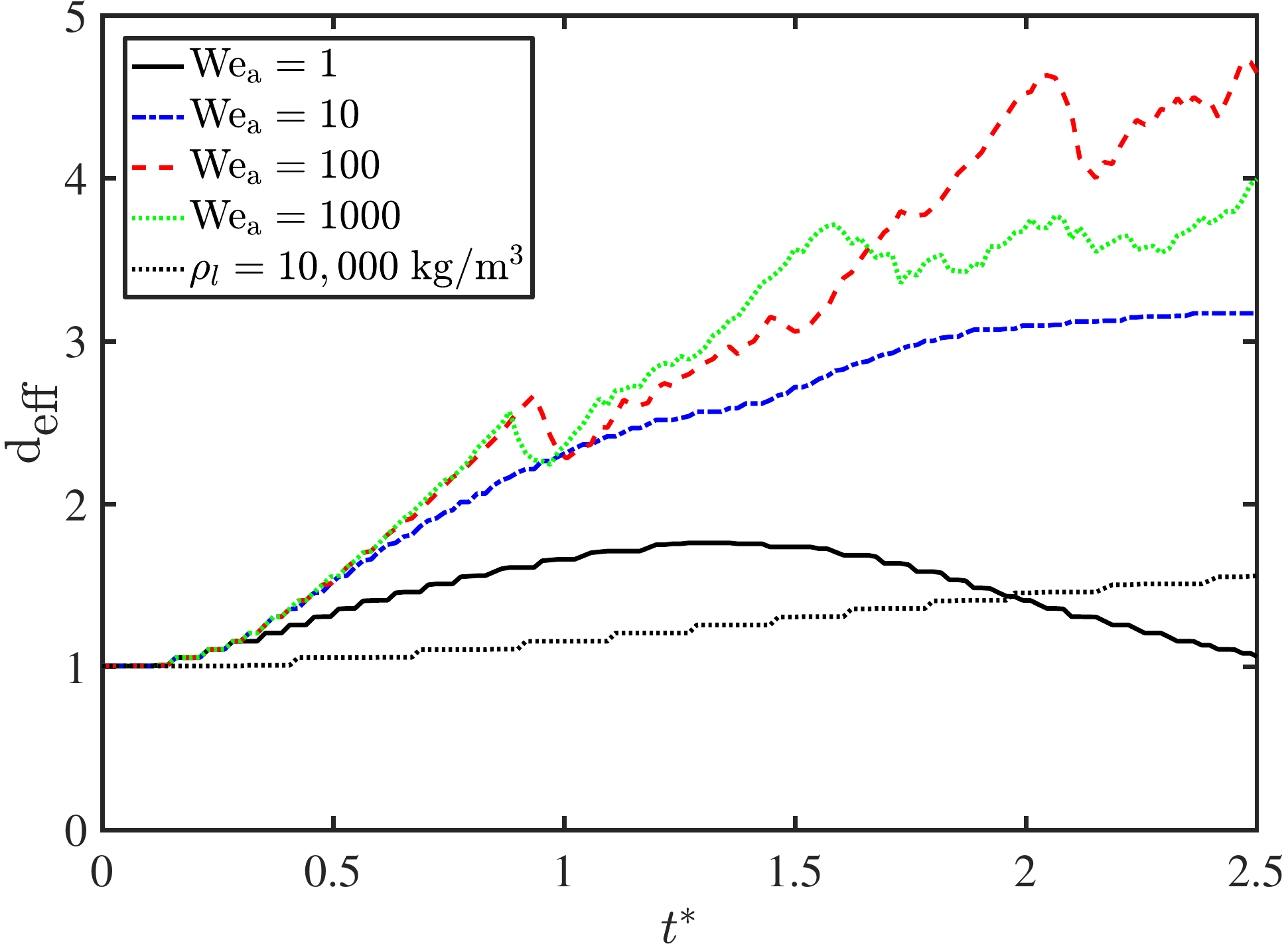}}\\
	\subfigure[$M_s=2.5$\label{fig:deff3}]{\includegraphics[width=0.45\textwidth]{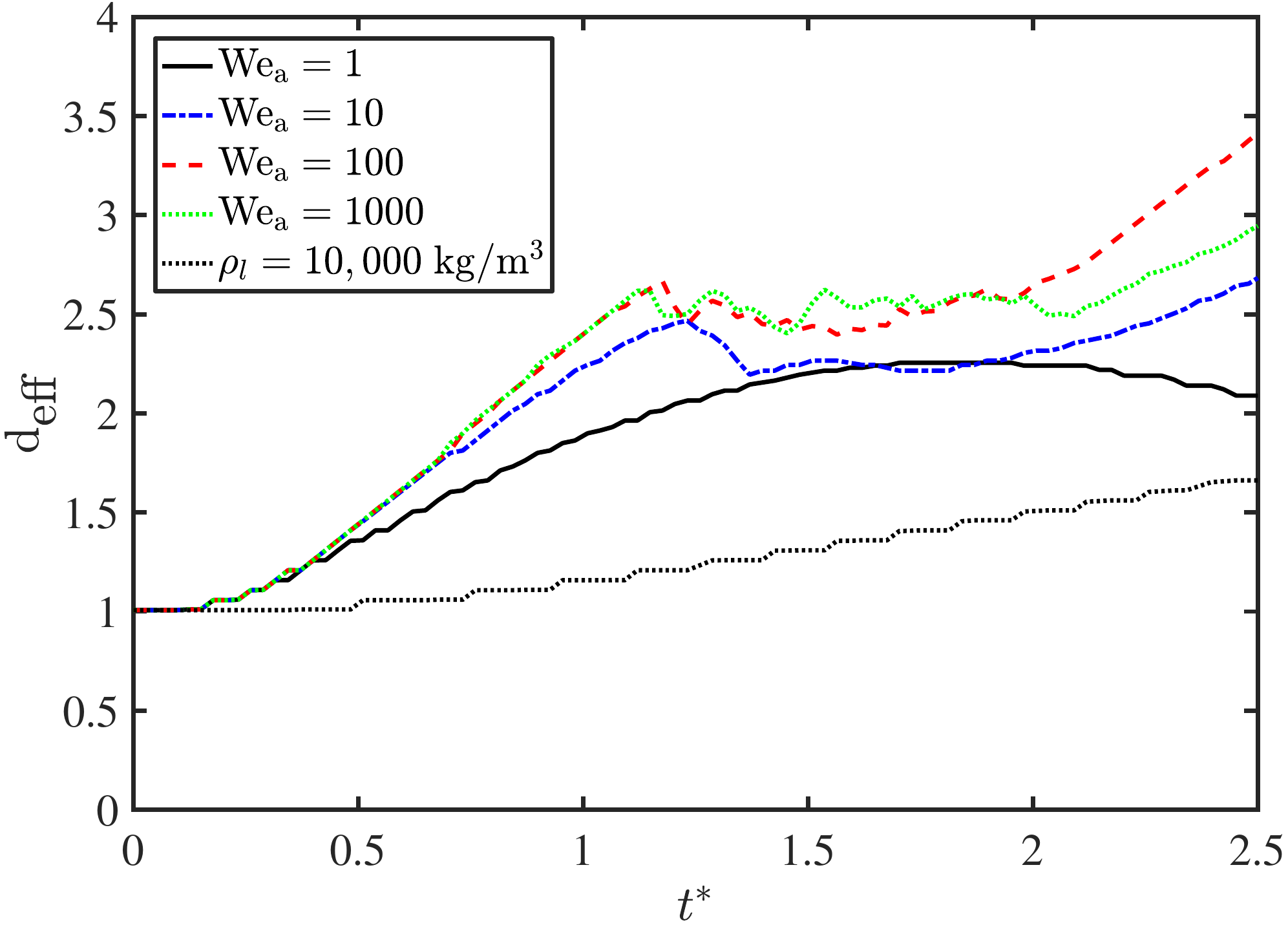}}
		\subfigure[$M_s=3$\label{fig:deff4}]{\includegraphics[width=0.45\textwidth]{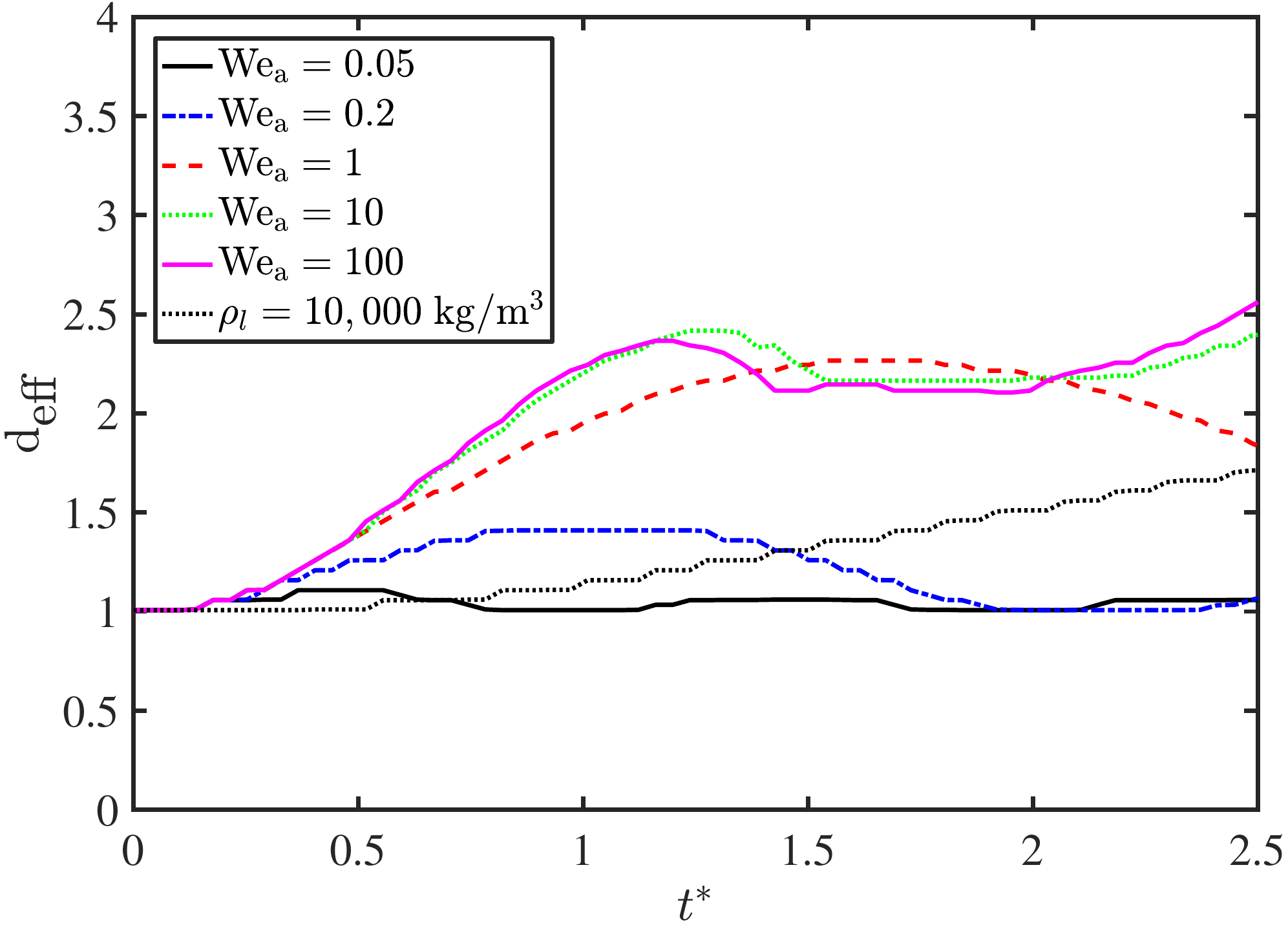}}
	\caption{\label{fig:deff} Effective deformed diameter comparison between acoustic Weber number at the four crossflow velocities.}
\end{figure}

To explore this, an effective diameter of the deformed drop was computed and the results are presented in Figure~\ref{fig:deff}.
This value is computed as the total projected length of the liquid on an x-normal plane, where the liquid is defined as $\phi>0.5$. 

Comparing the calculated effective diameters, a similar trend is observed for the effective diameter as the drag. 
Significant differences are seen in the effective diameter of the subsonic cases while less variation is observed in the supersonic cases at higher Weber numbers.
This suggests the similarities in drag are a product of a similar effective diameter throughout the breakup process, even if the breakup itself differs.

\begin{figure}\centering
	\subfigure[$M_s=1.47$\label{fig:cdp1}]{\includegraphics[width=0.45\textwidth]{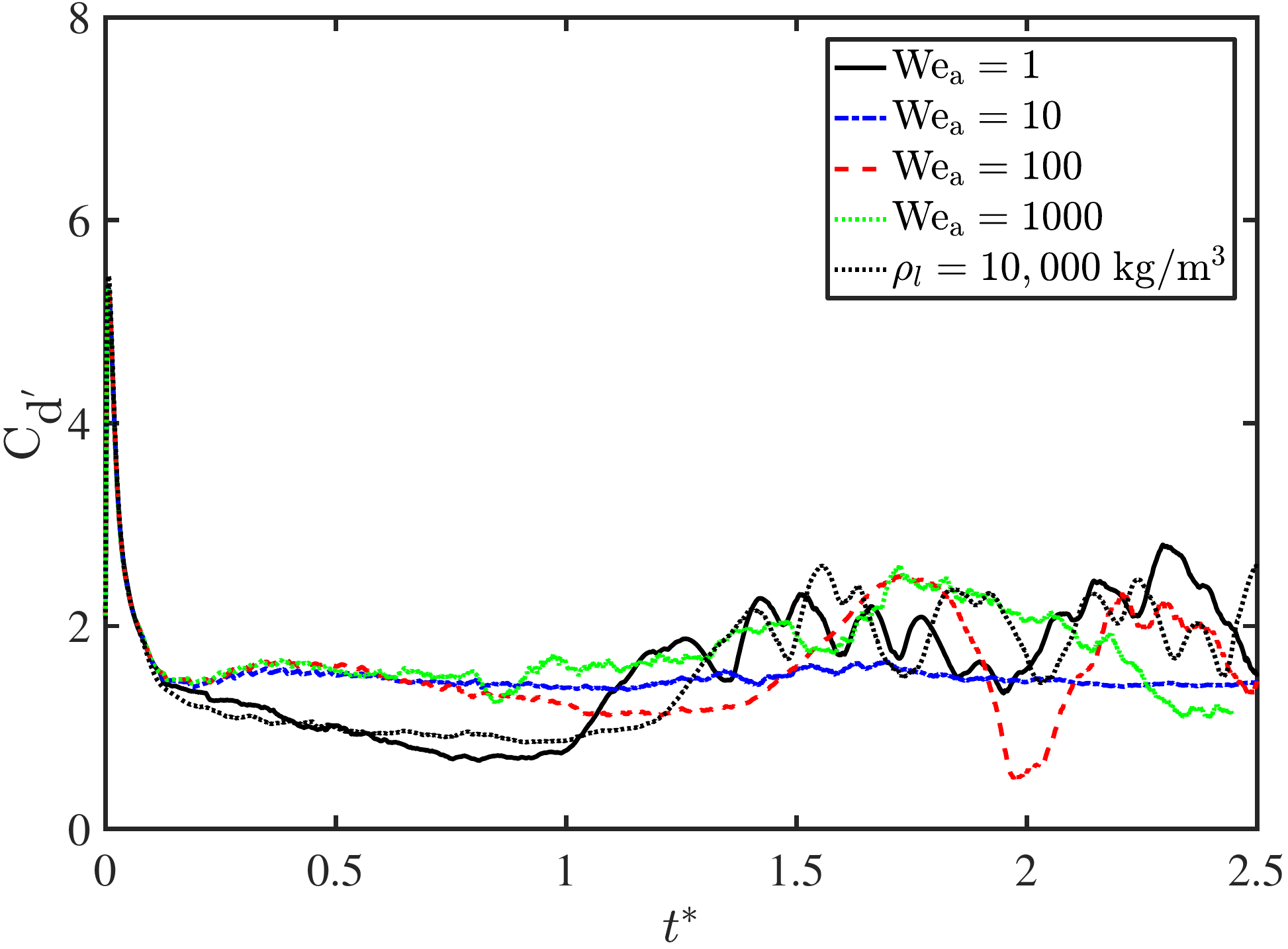}} 
	\subfigure[$M_s=2$\label{fig:cdp2}]{\includegraphics[width=0.45\textwidth]{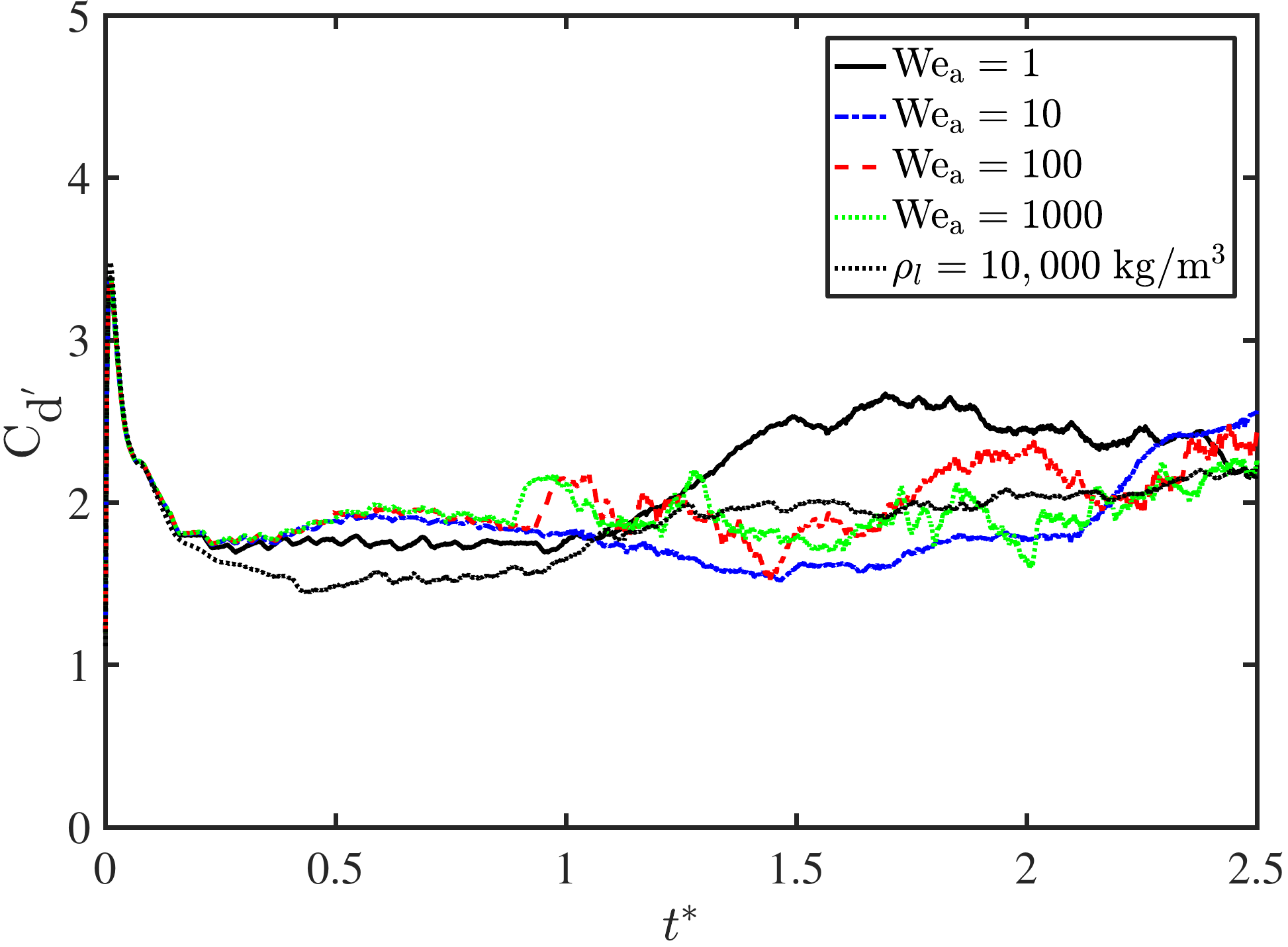}}\\
	\subfigure[$M_s=2.5$\label{fig:cdp3}]{\includegraphics[width=0.45\textwidth]{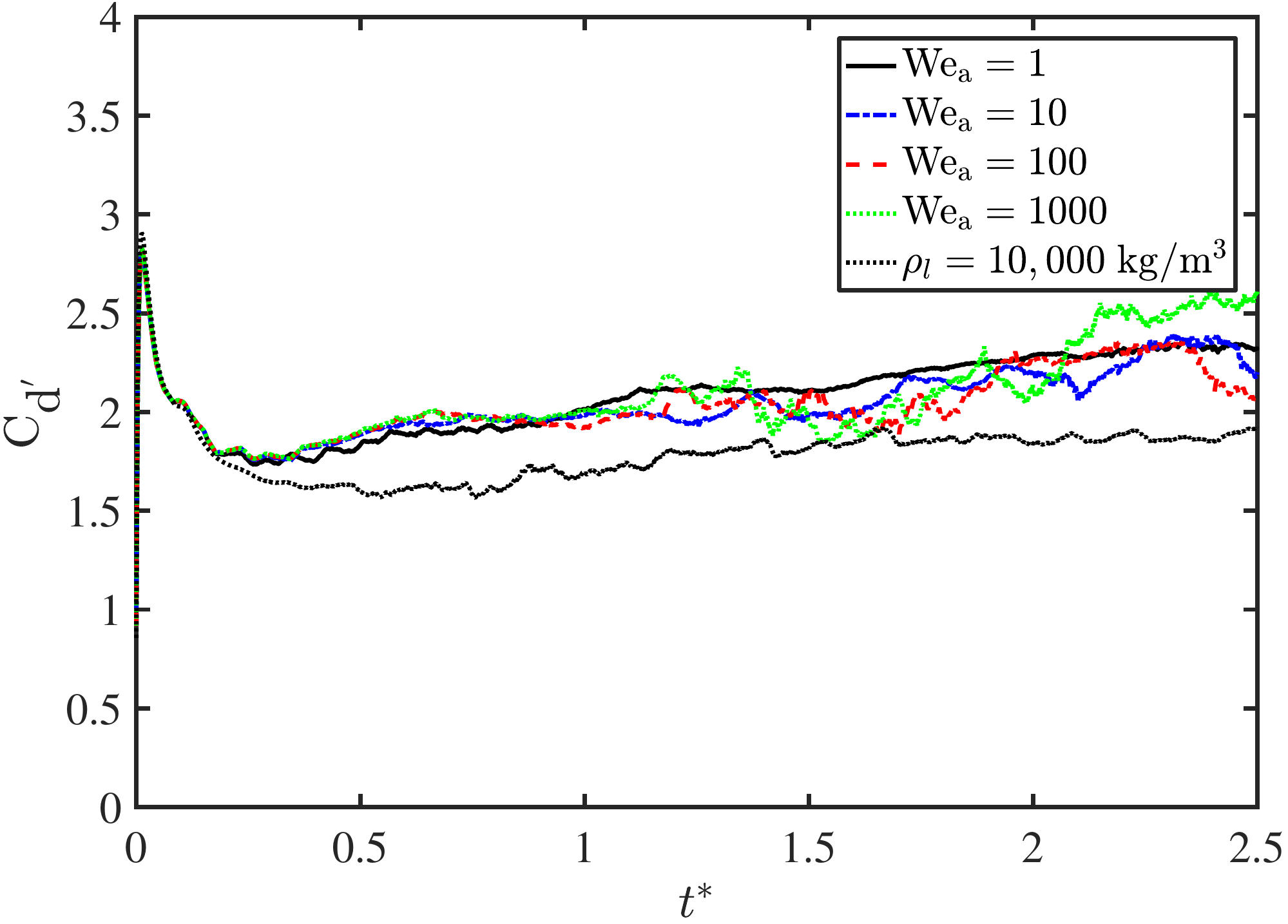}}
		\subfigure[$M_s=3$\label{fig:cdp4}]{\includegraphics[width=0.45\textwidth]{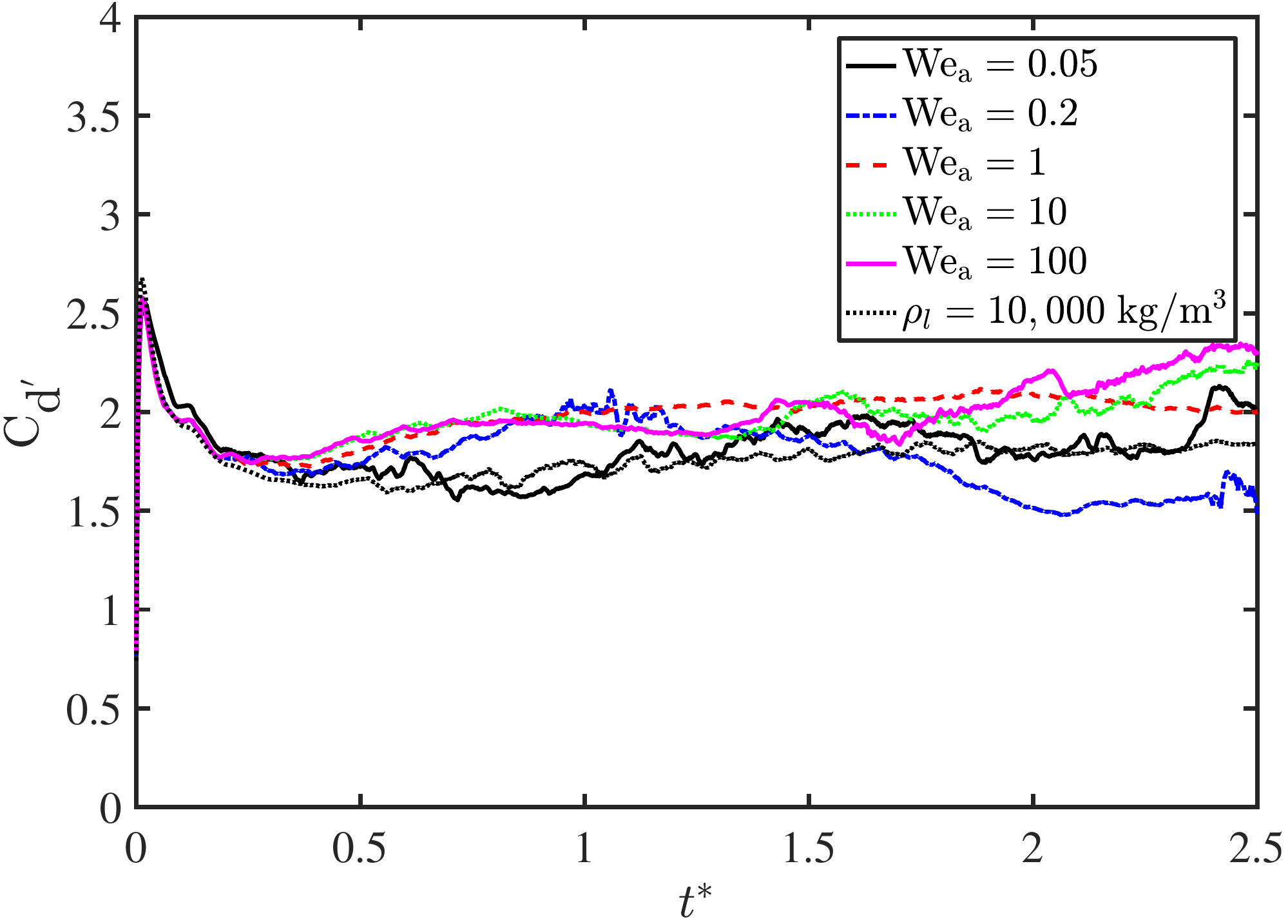}}
	\caption{\label{fig:cdp}Drag coefficient comparison using the effective diameter for calculation.}
\end{figure}

Figure~\ref{fig:cdp} depicts the drag coefficient computed again using Eq.~\ref{eqn:cd} but with the time dependent effective diameter used in place of the undeformed diameter term $d_0$. As noted by Meng and Colonius \cite{Meng2014}, the computed drag coefficients can largely be assumed as constant regardless of shock speed during the early stages of breakup when accounting for the effective deforming diameter of the droplets. Interestingly, the present simulations show that this assumption is still relatively reasonable during the mid and later stages of breakup and even when accounting for the effects of surface tension across a wide range of Weber numbers.  This is a notable result given the wide range of breakup behaviours observed in the present simulations.  These results are especially relevant at supersonic speeds where less variation of the drag coefficient is observed as a function of Weber number.

\section{Three-dimensional simulation of droplet breakup}\label{sec:3dsim}
\begin{figure}\centering
\subfigure[]{\includegraphics[width=0.3\textwidth]{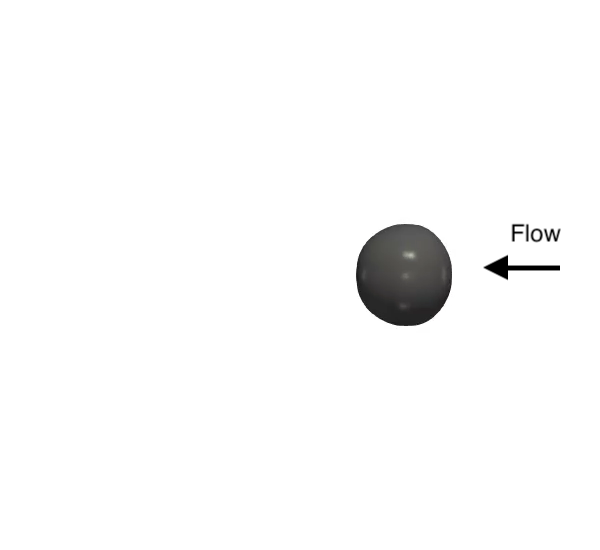}}
\subfigure[]{\includegraphics[width=0.3\textwidth]{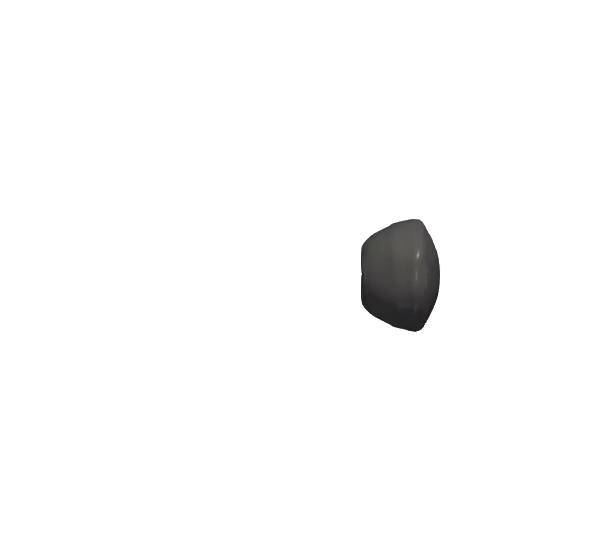}}
\subfigure[]{\includegraphics[width=0.3\textwidth]{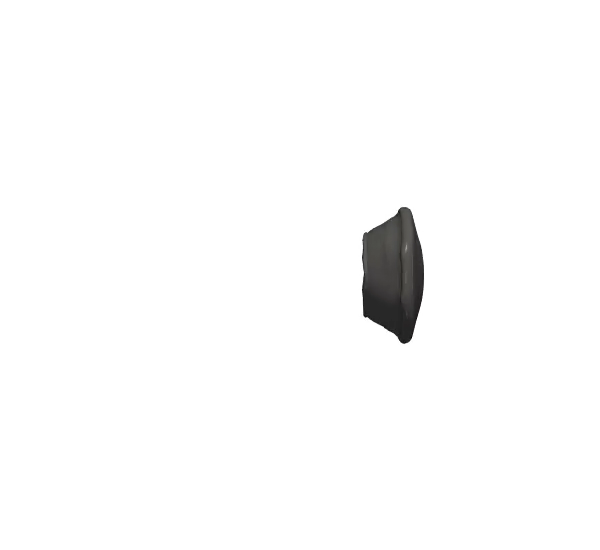}}
\subfigure[]{\includegraphics[width=0.3\textwidth]{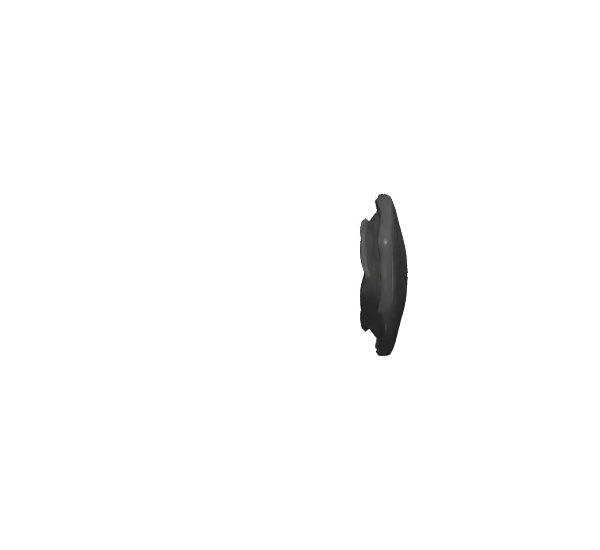}}
\subfigure[]{\includegraphics[width=0.3\textwidth]{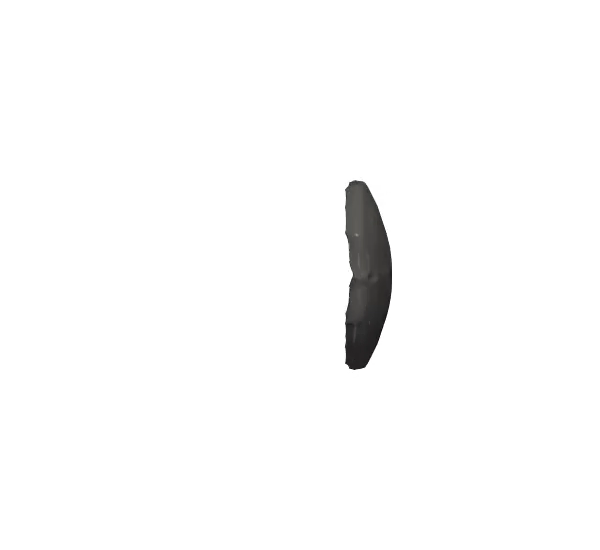}}
\subfigure[]{\includegraphics[width=0.3\textwidth]{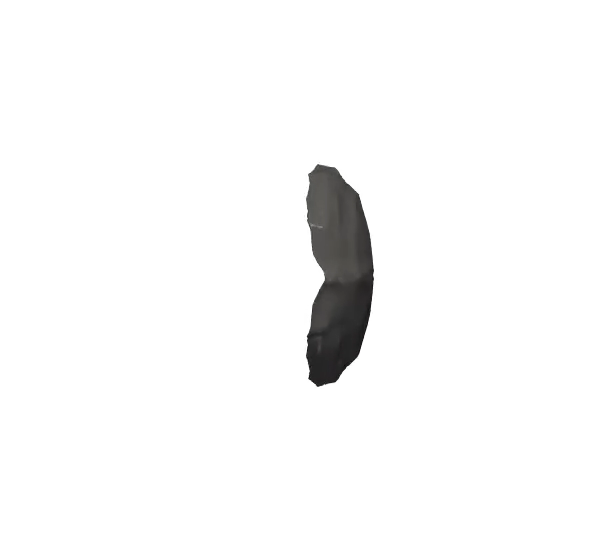}}
\subfigure[]{\includegraphics[width=0.3\textwidth]{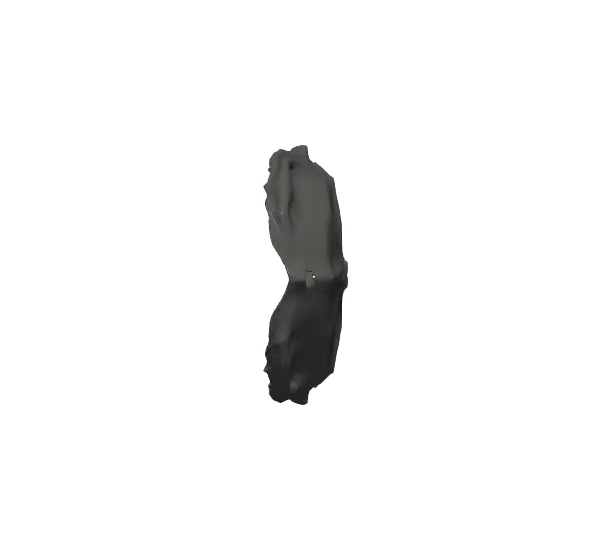}}
\subfigure[]{\includegraphics[width=0.3\textwidth]{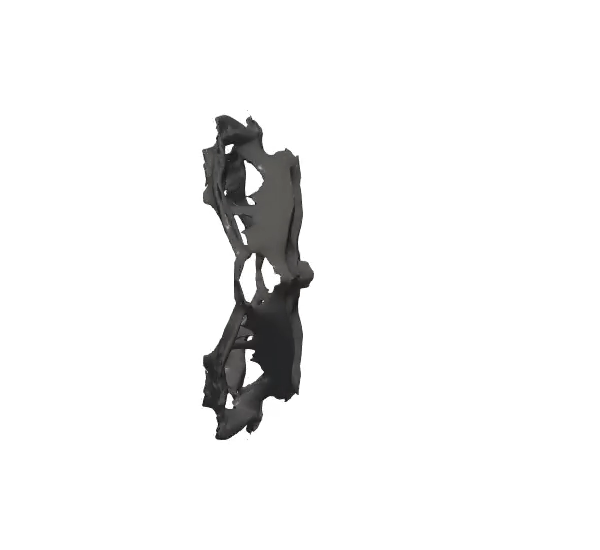}}
\subfigure[]{\includegraphics[width=0.3\textwidth]{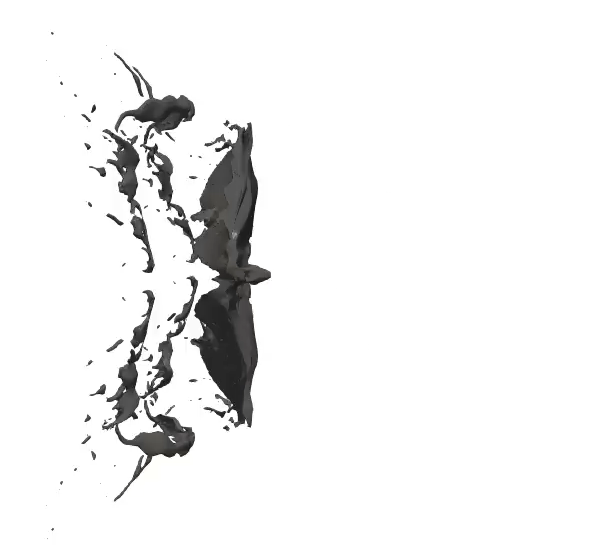}}
	\caption{\label{fig:3dsim} Snapshots from three-dimensional droplet breakup simulation corresponding to experiment from Figure 33 of Theofanous et al ~\cite{Theofanous2012}. Simulation consists of a water droplet impacted by a shock-wave with freestream flow conditions $M=0.32$, $\mathrm{Re}_{g}=2.2 \times 10^4$, $\mathrm{We}= 7.8 \times 10^2$, and $\mathrm{Oh}=2.4 \times 10^{-3}$. Note the crossflow is moving from right to left.}
\end{figure}

\revc{
A three-dimensional simulation of droplet breakup was performed. The objective being to further validate the ability of the numerical method to predict three-dimensional droplet breakup behaviour and to provide a point of comparison to the two-dimensional liquid column breakup simulations.

The flow conditions were set to match the experimental conditions in Figure 33 of Theofanous et al \cite{Theofanous2012}. Specifically, the simulation consists of a water droplet impacted by a shockwave with post-shock crossflow conditions of $M=0.32$, $\mathrm{Re}_{g}=2.2 \times 10^4$, $\mathrm{We}= 7.8 \times 10^2$, and $\mathrm{Oh}=2.4 \times 10^{-3}$.

Given the computational complexity of such a three-dimensional simulation, the grid resolution in the vicinity of the droplet was set to a relatively coarse $D/80$ and symmetry boundary conditions were employed at the centerline such that the computational domain consisted of only a quarter of the overall droplet. 

Figure~\ref{fig:3dsim} shows the progression of the droplet deformation and breakup. The present resolution is inadequate to capture the fine scale features of the breakup process however the overall droplet shape evolution over time reasonably agrees with the experimental behavior shown in the video supplementing Figure 33 of Theofanous et al \cite{Theofanous2012} (see supplementary multimedia material of \cite{Theofanous2012} for video).
}

\section{Conclusion}\label{sec:conclusions3}
Numerical experiments are performed of $M_s=1.47, 2, 2.5$ and $M_s=3$ shockwaves interacting with liquid columns at various Weber numbers.
The simulations account for the effects of compressibility, molecular viscosity, and surface tension.
The shockwaves induce a crossflow leading to aerobreakup of the liquid column.
A diverse range of complex interface dynamics and breakup modes are observed with good correlation to experimentally observed behavior across the range of Weber numbers tested.
During the early stages of the breakup process (i.e deformation), similar behavior is observed across the range of Mach numbers tested.
However, at later times the breakup behavior varies significantly depending on both the Mach and Weber numbers.
Additionally, lower Weber numbers result in lower observed drag coefficients for the liquid columns.
Depending on the Weber number, the drag coefficients are still approximately two to three times those observed for a rigid liquid column.
As a function of the Weber number, significantly less variation in the drag coefficient and qualitative flow features is observed as the Mach number increases.
In addition, when utilizing a deformed diameter in the drag coefficient calculation the results show significantly reduced variation between Weber numbers across all Mach numbers.
This has implications for subgrid atomization models which determine droplet trajectories based on estimated particle drag coefficients. 
\revc{A three-dimensional simulation, while under-resolved, displays reasonable agreement with the corresponding experimental breakup behavior, highlighting the potential of the numerical approach for future investigations. }

\section{Acknowledgments}
This work is supported by Taitech, Inc.
under sub-contracts TS15-16-02-004 and TS16-16-61-004 (primary contract FA8650-14-D-2316).
The computational resources in this paper are partially supported by the HPC@ISU equipment at Iowa State University, some of which has been purchased through funding provided by NSF under MRI grant number CNS 1229081 and CRI grant number 1205413. This work has been approved for unlimited release: LA-UR-19-25304.

\bibliography{ms}

\section{Appendix}
\begin{figure}
	\includegraphics[width=.40\textwidth,center]{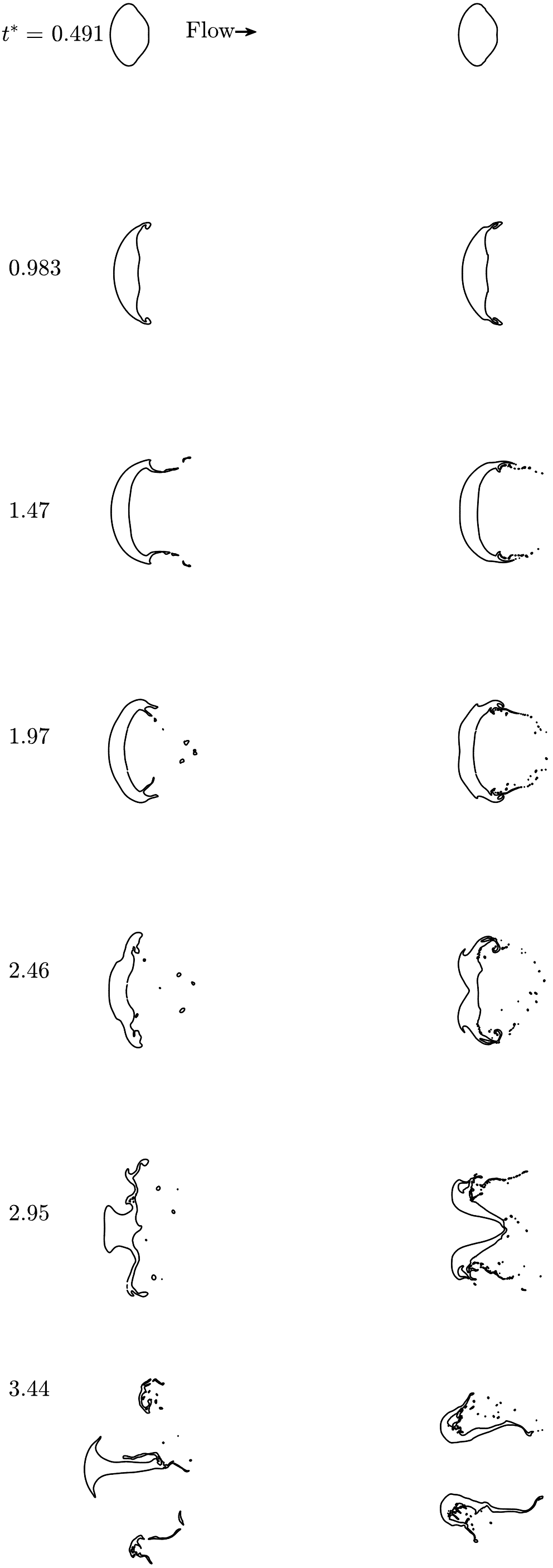} 
\begin{tabular*}{\textwidth}{lll @{\extracolsep{\fill}} ll}
		  &\hspace{50pt} (a) & \hspace{95pt} (b)       \\
		$\hspace{100pt}\mathrm{We_a}$ &\hspace{52pt} 20  &\hspace{91pt} 1000    \\
		$\hspace{100pt}\mathrm{We_c}$ & \hspace{52pt}457  & \hspace{90pt}22857 \\
		$\hspace{100pt}\mathrm{We_{eff}}$ & \hspace{52pt}283 &\hspace{90pt}14143 \\
	\end{tabular*}
	\caption{\label{fig:Ms3additional} Additional $M_s=3$ deformation and breakup behavior.}
\end{figure}

The additional cases for the $M_s=3.0$ incident shock are presented in Fig~\ref{fig:Ms3additional} for completeness.
Similar breakup characteristics are seen to those observed in Fig~\ref{fig:Ms3} in the comparable Weber number ranges.

\end{document}